\documentclass[longauth]{aa}  

\usepackage{graphicx}
\usepackage{amsmath}	% Advanced maths commands
\newcommand{\mstar}{\rm log(M_{star}/M_{\odot})}

\newcommand{\neIIIlam}{[\textrm{Ne}\textsc{III}]\ensuremath{\lambda}3869}
\newcommand{\nevlam}{[\textrm{Ne}\textsc{V}]\ensuremath{\lambda}3427}

\newcommand{\oiilam}{[\textrm{O}\textsc{II}]\ensuremath{\lambda\lambda}3727,3729}

\usepackage[colorlinks=true,allcolors=blue]{hyperref}
\usepackage{orcidlink}
\usepackage{comment}

\usepackage{txfonts}
\begin{document} 

\title{Beyond Traditional Diagnostics: Identifying Active Galactic Nuclei with Spectral Energy Distribution Fitting in DESI Data}
\titlerunning{DESI SED-selected AGN}

\author{M.~Siudek\orcidlink{0000-0002-2949-2155}\thanks{\email{msiudek@iac.es}}\inst{\ref{aff1},\ref{aff2}}
\and 
M.~Mezcua\orcidlink{0000-0003-4440-259X}\inst{\ref{aff2},\ref{aff3}}
\and
C.~Circosta\inst{\ref{aff4}}
\and
C.~Maraston\inst{\ref{aff5}}
\and
J.~Moustakas\orcidlink{0000-0002-2733-4559}\inst{\ref{aff6}}
\and
H.~Zou\orcidlink{0000-0002-6684-3997}\inst{\ref{aff7}}
\and
J.~Aguilar\inst{\ref{aff8}}
\and
S.~Ahlen\orcidlink{0000-0001-6098-7247}\inst{\ref{aff9}}
\and
D.~Bianchi\orcidlink{0000-0001-9712-0006}\inst{\ref{aff10},\ref{aff11}}
\and
D.~Brooks\inst{\ref{aff4}}
\and
T.~Claybaugh\inst{\ref{aff8}}
\and
K.~S.~Dawson\orcidlink{0000-0002-0553-3805}\inst{\ref{aff12}}
\and
A.~de la Macorra\orcidlink{0000-0002-1769-1640}\inst{\ref{aff13}}
\and
Arjun~Dey\orcidlink{0000-0002-4928-4003}\inst{\ref{aff14}}
\and
P.~Doel\inst{\ref{aff4}}
\and
J.~E.~Forero-Romero\orcidlink{0000-0002-2890-3725}\inst{\ref{aff15},\ref{aff16}}
\and
E.~Gaztañaga\inst{\ref{aff3},\ref{aff5},\ref{aff2}}
\and
S.~Gontcho A Gontcho\orcidlink{0000-0003-3142-233X}\inst{\ref{aff8}}
\and
G.~Gutierrez\inst{\ref{aff17}}
\and
M.~Ishak\orcidlink{0000-0002-6024-466X}\inst{\ref{aff18}}
\and
S.~Juneau\orcidlink{0000-0002-0000-2394}\inst{\ref{aff14}}
\and
D.~Kirkby\orcidlink{0000-0002-8828-5463}\inst{\ref{aff19}}
\and
T.~Kisner\orcidlink{0000-0003-3510-7134}\inst{\ref{aff8}}
\and
A.~Kremin\orcidlink{0000-0001-6356-7424}\inst{\ref{aff8}}
\and
A.~Lambert\inst{\ref{aff8}}
\and
M.~Landriau\orcidlink{0000-0003-1838-8528}\inst{\ref{aff8}}
\and
L.~Le~Guillou\orcidlink{0000-0001-7178-8868}\inst{\ref{aff20}}
\and
A.~Meisner\orcidlink{0000-0002-1125-7384}\inst{\ref{aff14}}
\and
R.~Miquel\inst{\ref{aff21},\ref{aff22}}
\and
F.~Prada\orcidlink{0000-0001-7145-8674}\inst{\ref{aff23}}
\and
I.~P\'erez-R\`afols\orcidlink{0000-0001-6979-0125}\inst{\ref{aff24}}
\and
G.~Rossi\inst{\ref{aff25}}
\and
E.~Sanchez\orcidlink{0000-0002-9646-8198}\inst{\ref{aff26}}
\and
D.~Schlegel\inst{\ref{aff8}}
\and
M.~Schubnell\inst{\ref{aff27},\ref{aff28}}
\and
H.~Seo\orcidlink{0000-0002-6588-3508}\inst{\ref{aff29}}
\and
D.~Sprayberry\inst{\ref{aff14}}
\and
G.~Tarl\'{e}\orcidlink{0000-0003-1704-0781}\inst{\ref{aff28}}
\and
B.~A.~Weaver\inst{\ref{aff14}}
}

\institute{Instituto de Astrof\'isica de Canarias (IAC); Departamento de Astrof\'isica, Universidad de La Laguna (ULL), 38200, La Laguna, Tenerife, Spain\label{aff1}
\and
Institute of Space Sciences (ICE, CSIC), Campus UAB, Carrer de Can Magrans, s/n, 08193 Barcelona, Spain\label{aff2}
\and
Institut d'Estudis Espacials de Catalunya (IEEC), c/ Esteve Terradas 1, Edifici RDIT, Campus PMT-UPC, 08860 Castelldefels, Spain\label{aff3}
\and
Department of Physics \& Astronomy, University College London, Gower Street, London, WC1E 6BT, UK\label{aff4}
\and
Institute of Cosmology and Gravitation, University of Portsmouth, Dennis Sciama Building, Portsmouth, PO1 3FX, UK\label{aff5}
\and
Department of Physics and Astronomy, Siena College, 515 Loudon Road, Loudonville, NY 12211, USA\label{aff6}
\and
National Astronomical Observatories, Chinese Academy of Sciences, A20 Datun Road, Chaoyang District, Beijing, 100101, P.~R.~China\label{aff7}
\and
Lawrence Berkeley National Laboratory, 1 Cyclotron Road, Berkeley, CA 94720, USA\label{aff8}
\and
Department of Physics, Boston University, 590 Commonwealth Avenue, Boston, MA 02215 USA\label{aff9}
\and
Dipartimento di Fisica ``Aldo Pontremoli'', Universit\`a degli Studi di Milano, Via Celoria 16, I-20133 Milano, Italy\label{aff10}
\and
INAF-Osservatorio Astronomico di Brera, Via Brera 28, 20122 Milano, Italy\label{aff11}
\and
Department of Physics and Astronomy, The University of Utah, 115 South 1400 East, Salt Lake City, UT 84112, USA\label{aff12}
\and
Instituto de F\'{\i}sica, Universidad Nacional Aut\'{o}noma de M\'{e}xico,  Circuito de la Investigaci\'{o}n Cient\'{\i}fica, Ciudad Universitaria, Cd. de M\'{e}xico  C.~P.~04510,  M\'{e}xico\label{aff13}
\and
NSF NOIRLab, 950 N. Cherry Ave., Tucson, AZ 85719, USA\label{aff14}
\and
Departamento de F\'isica, Universidad de los Andes, Cra. 1 No. 18A-10, Edificio Ip, CP 111711, Bogot\'a, Colombia\label{aff15}
\and
Observatorio Astron\'omico, Universidad de los Andes, Cra. 1 No. 18A-10, Edificio H, CP 111711 Bogot\'a, Colombia\label{aff16}
\and
Fermi National Accelerator Laboratory, PO Box 500, Batavia, IL 60510, USA\label{aff17}
\and
Department of Physics, The University of Texas at Dallas, 800 W. Campbell Rd., Richardson, TX 75080, USA\label{aff18}
\and
Department of Physics and Astronomy, University of California, Irvine, 92697, USA\label{aff19}
\and
Sorbonne Universit\'{e}, CNRS/IN2P3, Laboratoire de Physique Nucl\'{e}aire et de Hautes Energies (LPNHE), FR-75005 Paris, France\label{aff20}
\and
Instituci\'{o} Catalana de Recerca i Estudis Avan\c{c}ats, Passeig de Llu\'{\i}s Companys, 23, 08010 Barcelona, Spain\label{aff21}
\and
Institut de F\'{i}sica d’Altes Energies (IFAE), The Barcelona Institute of Science and Technology, Edifici Cn, Campus UAB, 08193, Bellaterra (Barcelona), Spain\label{aff22}
\and
Instituto de Astrof\'{i}sica de Andaluc\'{i}a (CSIC), Glorieta de la Astronom\'{i}a, s/n, E-18008 Granada, Spain\label{aff23}
\and
Departament de F\'isica, EEBE, Universitat Polit\`ecnica de Catalunya, c/Eduard Maristany 10, 08930 Barcelona, Spain\label{aff24}
\and
Department of Physics and Astronomy, Sejong University, 209 Neungdong-ro, Gwangjin-gu, Seoul 05006, Republic of Korea\label{aff25}
\and
CIEMAT, Avenida Complutense 40, E-28040 Madrid, Spain\label{aff26}
\and
Department of Physics, University of Michigan, 450 Church Street, Ann Arbor, MI 48109, USA\label{aff27}
\and
University of Michigan, 500 S. State Street, Ann Arbor, MI 48109, USA\label{aff28}
\and
Department of Physics \& Astronomy, Ohio University, 139 University Terrace, Athens, OH 45701, USA\label{aff29}
}

   \date{Received xxx; accepted xxx}

% \abstract{}{}{}{}{} 
% 5 {} token are mandatory

  \abstract
   {}
   {Active galactic nuclei (AGN) are typically identified through their distinctive X-ray or radio emissions, mid-infrared (MIR) colors, or emission lines. However, each method captures different subsets of AGN due to signal-to-noise (SNR) limitations, redshift coverage, and extinction effects, underscoring the necessity for a multi-wavelength approach for comprehensive AGN samples. This study explores the effectiveness of spectral energy distribution (SED) fitting as a robust method for AGN identification.}
  {Using {\tt CIGALE} optical-MIR SED fits on DESI Early Data Release galaxies, we compare SED-based AGN selection ({\tt AGNFRAC} $\geq0.1$) with traditional methods including BPT diagrams, WISE colors, X-ray, and radio diagnostics.}
  { SED fitting identifies $\sim 70\%$ of narrow/broad-line AGN and 87\% of WISE-selected AGN. Incorporating high SNR WISE photometry reduces star-forming galaxy contamination from 62\% to 15\%. Initially, $\sim50\%$ of SED-AGN candidates are undetected by standard methods, but additional diagnostics classify $\sim85\%$ of these sources, revealing LINERs and retired galaxies potentially representing evolved systems with weak AGN activity. Further spectroscopic and multi-wavelength analysis will be essential to determine the true AGN nature of these sources. 
   }
  {SED fitting provides complementary AGN identification, unifying multi-wavelength AGN selections. This approach enables more complete -- albeit with some contamination -- AGN samples essential for upcoming large-scale surveys where spectroscopic diagnostics may be limited.}


   \maketitle
%
%%%%%%%%%%%%%%%%% BODY OF PAPER %%%%%%%%%%%%%%%%%%

\section{Introduction}\label{sec:introduction}

Supermassive black holes with masses of more than a million times that of the Sun~\citep[e.g.][]{Kormendy2013}, are believed to inhabit the centers of galaxies, including the Milky Way~\citep[e.g.,][]{Ghez2008, Genzel2010a}. 
Despite the key role supermassive black holes are thought to play in galaxy formation and evolution, their origin and impact on host galaxies are still a major challenge of modern astrophysics. 
When supermassive black holes are actively accreting as active galactic nuclei (AGN), they are brighter and easier to detect and thus serve as a proxy to study supermassive black hole properties.
Thus, identifying AGN is crucial for understanding galaxy evolution, black hole growth, and cosmic feedback processes (see \citealt{Harrison2024} for a review). 

Despite their important role, a uniform method to identify AGN is challenging as different selection techniques lead to different samples that may not even overlap with each other. 
AGN are usually identified based on: i) their optical emission lines (e.g., \citealt{Baldwin1981, Kewley2001, Kauffmann2003}), and variability (e.g., \citealt{Pai2024}) ii) X-ray emission (e.g., \citealt{Brandt2015}, iii) mid-infrared (MIR) emission (e.g., \citealt{Lacy2004, Stern2005, Jarrett2011, Stern2012, Assef2013, Lacy2015, Hviding2022}),  and variability (e.g., \citealt{Bernal2025}), and iv) radio emission
(e.g., \citealt{Heckman2014, Padovani2016, Tadhunter2016}). 
Optical spectroscopic selection identifies AGN by detecting broad (BL-AGN, i.e, type 1 AGN) or narrow (NL-AGN, i.e., type 2 AGN) emission lines, indicative of high-energy processes near the black hole~\citep[e.g.,][]{Baldwin1981}. However, star formation elevated by a diffuse ionized gas, photoionization by hot stars, metallicity (as indicated by strong forbidden lines such as [N II]$\lambda$6583, [S II]$\lambda$6716, 6731, and [O III]$\lambda$5007), or shocks can be misclassified as  AGN~\citep[e.g.,][]{Wylezalek2018}. 
X-ray selection is highly effective because AGN typically exhibit strong X-ray emission that is less likely to be contaminated by star formation processes. However, X-ray surveys are often limited by sensitivity and coverage and may miss obscured AGN~\citep[e.g.,][]{Gilli2007, Burlon2011, Mazzolari2024}. 
MIR selection uses color criteria to distinguish AGN from star-forming galaxies due to the characteristic dust emission heated by the AGN~\citep{Stern2005, Assef2013}.  The high-energy radiation emitted from the accretion disk heats up the surrounding dust, causing it to emit infrared radiation, particularly in the MIR regime ($5 - 30 \mu m$). The MIR emission has already proven its usefulness in AGN selection, though it is biased towards the selection of AGN that dominate over the emission from the host galaxies. 
Radio selection relies on the strong synchrotron emission from relativistic jets, but this method is biased towards radio-loud AGN, which are a minority of the AGN population. Moreover, radio selection can pick up star-forming regions with strong radio emission~\citep{Padovani2016}. 
Summarizing, accurate AGN identification is often challenging,
as each of these common methods captures different subsets of the AGN population and is subject to its own limitations and biases.

The aforementioned limitations of traditional selection methods highlight the need for a multi-wavelength approach to reliably and comprehensively identify AGN. 
The advent of large multi-wavelength surveys triggered using  SED fitting to constrain AGN and their host galaxy properties for statistical samples. 
Recently, the SED fitting approach revealed the potential not only to derive reliable properties of AGN and host properties (e.g., \citealt{Marshall2022, Mountrichas2021a, Burke2022, Best2023}) but also to identify AGN based on their multi-wavelength information (e.g., \citealt{Thorne2022, Best2023, Yang2023, Prathap2024}). AGN SED modeling techniques are also used as the base for the target selection of forthcoming wide-field spectroscopic surveys such as 4MOST (\citealt{Merloni2019}) and VLT-MOONS (\citealt{Maiolino2020}). 

In this paper, we identify AGN in the DESI survey using the {\tt CIGALE} SED fitting code, modeling optical to MIR photometry for millions of sources across diverse types and redshifts. 
In particular, DESI has already revealed unprecedented samples of dust-reddened QSO~\citep{Fawcett2023} or changing-look AGN~\citep{Guo2024, Guo2024b, Guo2025}. DESI value-added catalog (VAC) of AGN selected based on their multi-wavelength information (Juneau et al. in prep.) provides so far the largest AGN spectroscopic sample\footnote{\url{https://data.desi.lbl.gov/doc/releases/dr1/vac/agnqso/}}. This will allow for a more detailed study of the role of AGN in galaxy evolution. 

Our study explores the potential of SED-based AGN identification and assesses the necessity of MIR data in minimizing the misclassification of star-forming galaxies as AGN. By integrating multi-wavelength data, we aim to contribute to a more accurate and complete catalog of AGN in the DESI survey.  
The structure of the paper is as follows. In Sect. \ref{sec:data}, we provide an overview of the DESI data. In Sect. \ref{sec:AGNidentification}, we describe the AGN selection using different standard techniques. Results with discussion are presented in Sect.~\ref{sec:resultsanddiscussion}. Finally, Sect. \ref{sec:conclusions} summarizes the efficiency of SED-based AGN identification.

Throughout this paper, we assume WMAP7 cosmology \citep{Komatsu2011}, with $\Omega_{m}$ = 0.272 and $H_{0}$ = 70.4. We also consider the photometry in AB magnitudes (\citealt{Oke1983}).

\section{DESI data}

\subsection{Overview of DESI}\label{sec:data}

DESI is a robotic, 5000-fiber multiobject spectroscopic surveyor operating on the Mayall 4-meter telescope at Kitt Peak National Observatory~\citep{DESICollaboration2022}. DESI is capable of obtaining simultaneous spectra covering a range of $3\,600-9\,800$ {\rm \AA}  with a resolution of $\rm R = 2\,000 - 5\,500$ of almost 5\,000 objects over a $\sim 3 \deg^2$ field~\citep{DESICollaboration2016, DESI2016b.Instr, Corrector.Miller.2023, FiberSystem.Poppett.2024}. DESI is currently conducting a five-year survey to observe approximately 36 million galaxies~\citep{Hahn2023a, Raichoor2023, Zhou2023} and 3 million quasars~\citep{Chaussidon2023} over 14\,000 $\rm deg^2$~\citep{DESICollaboration2023, DESIDr12025}. This campaign will provide ten times more spectra than the  SDSS~\citep{York2000, Almeida2023} sample of extragalactic targets and substantially deeper than prior large-area surveys~\citep{DESICollaboration2023}. 
With the aim of determining the nature of dark energy, DESI will provide the most precise measurement of the expansion history of the universe ever obtained~\citep{Levi2013}. 

DESI started to operate in December 2020, with a 5-month Survey Validation~\citep[SV;][]{DESI2023a.KP1.SV} before the start of the Main Survey. The entire SV data, internally known as {\it Fuji}, is publicly released as the DESI Early Data Release~\citep[EDR;][]{DESICollaboration2023} while the First Data Release~\citep[DR1;][]{DESIDr12025} was released in March 2025. This work relies on the EDR data as the complementary studies of the VAC presented in~\citet{Siudek2024}. The expansion to DR1 is left for future work. The DR1 already showcases the DESI potential by the cosmological results from the full-shape analysis~\citep{DESI2024.VII.KP7B}. 

The scale of DESI observations is supported by software pipelines and products, which include imaging from the DESI Legacy Imaging Surveys~\citep{Zou2017, Dey2019}, a fully automatic spectroscopic reduction pipeline~\citep{Guy2023, SurveyOps.Schlafly.2023}, followed by a template-fitting pipeline to derive classifications and redshifts for each targeted source ({\tt Redrock}\footnote{\url{https://github.com/desihub/redrock}}; \citealt{Anand2024}; Bailey et al. in prep.), and for the special case of QSOs \citep{Brodzeller2023}. {\tt Redrock} provides the redshift ({\tt Z}), redshift uncertainty ({\tt ZERR}), a redshift warning bitmask ({\tt ZWARN}), and a spectral type ({\tt SPECTYPE}) to every target based on the best fit. 
The photometry in $g$, $r$, and $z$ bands is estimated from the Legacy survey (DR9) images with the {\tt Tractor} \footnote{\url{https://github.com/dstndstn/tractor}} inference modeling code~\citep{Lang2016}. Mid-infrared photometry is derived via forced photometry of the unWISE co-adds based on this optical model~\citep{Meisner2021}.

\subsection{DESI EDR VAC of physical properties}\label{sec:DesiEDR}

The DESI EDR redshift catalog consists of 2\,847\,435 sources~\citep{DESICollaboration2023}. For a sample of 1\,337\,250 galaxies and quasars (with {\tt SPECTYPE = GALAXY | QSO}) characterized by reliable redshift  ({\tt ZWARN = 0 or 4} 
 and that do not have any fiber issues ({\tt COADD\_FIBERSTATUS = 0}) 
 we created a VAC of physical properties, including stellar masses and star formation rates (SFRs) as well as AGN features~\citep{Siudek2024}. 
Here, we present a short summary of the VAC and refer the reader to \cite{Siudek2024} for the full description.  

Physical properties of DESI galaxies are derived by performing SED fitting using Code Investigating GALaxy Emission~\citep[{\tt CIGALE} v2022.1;][]{Boquien2019} relying on the optical-MIR photometry and spectroscopic redshifts. 
{\tt CIGALE} already proved its efficiency in deriving physical properties of galaxies hosting AGN (e.g., \citealt{Salim2016, Malek2018, Osborne2024, Csizi2024}),  as well as high-z AGN (e.g., \citealt{Juodzbalis2023, Yang2023, Burke2024, Mezcua2023, Mezcua2024}). 
{\tt CIGALE} uses a library of models to fit the SED of galaxies, where the best fit is found based on the reduced $\chi^2$ minimization among all the possible combinations of the SED models. 
The library of models is created under the assumption of a delayed SFH with an optional exponential burst~\citep{Ciesla2015} and the~\cite{Bruzual2003} SSP models adopting a \cite{Chabrier2003} initial mass function.  We assume solar metallicity and use \cite{Fritz2006} models to account for possible AGN contribution. 
{ The grid of input parameters used to generate the model library follows the configuration adopted in the DESI EDR VAC~\citep{Siudek2024}, and is summarized in Table~\ref{tab:SEDParameters} in  Appendix~\ref{app:CIGALETable}. }
 The generated models cover a wide range of objects, including galaxies without AGN contribution, as well as BL-, and NL-AGN. This provides flexibility and allows us to build the catalog of physical properties for both AGN and non-AGN host galaxies.  

The SEDs of observed galaxies are created from the ground-based optical and near-infrared (NIR) photometry (i.e., {\it grz} photometry, which we shortly refer to as optical) complemented by observations from MIR bands at $\rm 3.4, 4.6, 12$ and 22 $\mu$m provided by the Wide-field Infrared Survey Explorer (WISE;~\citealt{Wright2010}), a mission extension NEOWISE-Reactivation forced-photometry~\citep{Mainzer2014}.

We note that the precise AGN classification of individual sources -- especially those with low AGN fractions -- as well as the stellar masses and SFRs may depend on the model assumptions, and the incorporation of the WISE photometry (see~\citealt{Siudek2024} for a discussion). 
Based on the representative sample of $\sim50,000$ galaxies including passive, star-forming, and AGN galaxies, ~\cite{Siudek2024} find that without WISE photometry the AGN fraction is overestimated ({\tt AGNFRAC $\ge 0.1$}) for 68\% of star-forming galaxies selected based on the [NII]-BPT emission line diagnostic diagram (see details in Sect.~\ref{sec:BPTdiagram}). The requirement of a high SNR in the WISE observations (forced by {\tt FLAGINFRARED} = 4)  reduces it to 19\%. This suggests that the MIR information is crucial to properly identify AGN via SED fitting. ~\cite{Siudek2024} also show that the overestimation of the {\tt AGNFRAC} for the star-forming galaxies without WISE information has a negligible effect on their stellar mass or SFR estimates. 
We further discuss the dependence of the AGN identification on the model choice in Appendix~\ref{app:model_dependence}.

\begin{figure*}
 	\centerline{\includegraphics[width=0.99\textwidth]{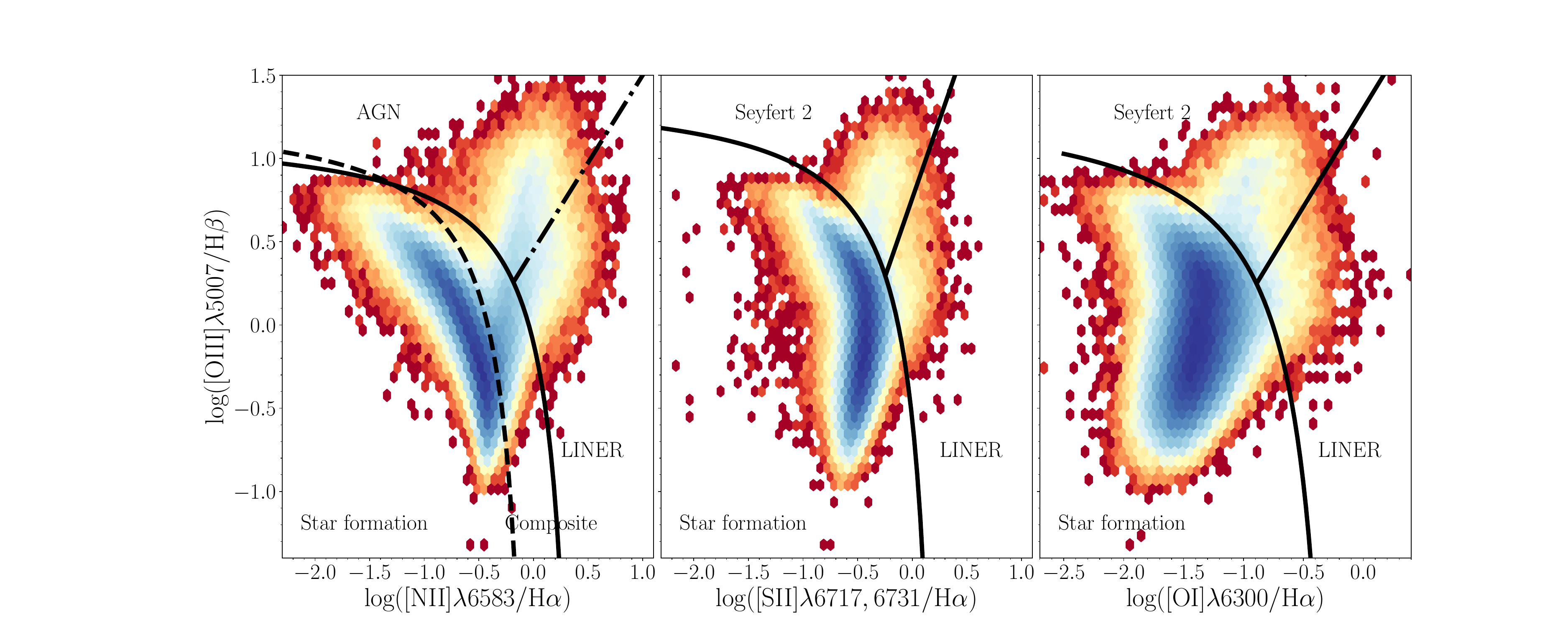}}
	\caption{Emission line diagnostic diagrams for DESI galaxies: N[II]-BPT (left), [SII]-BPT (middle), and [OI]-BPT (right). The demarcation lines separate star-forming galaxies, composites, and AGN (for simplicity we refer to Seyfert 2 galaxies as AGN). }
	\label{fig:BPT}
\end{figure*}  

\section{Sample selection based on the AGN diagnostics}\label{sec:AGNidentification}

{ In this paper, we rely on the sample of 510\,938 DESI EDR galaxies at $z \leq 0.5$ from the VAC of physical properties\footnote{\url{https://data.desi.lbl.gov/doc/releases/edr/vac/cigale/}}. We remove sources with poor photometric quality, defined as those with {\tt LOGM} $= 0$, and sources with poor SED fits, defined as those with {\tt CHI2} $> 17$ (see~\citealt{Siudek2024}). These criteria, along with the redshift limit ({\tt Z} $\leq 0.5$), define our parent sample. No additional cuts are applied at this stage; further selection criteria (e.g., SNR for emission-line quality or WISE-based selections) are introduced in the relevant sections.}

To validate the performance of the SED-based AGN identification in recovering true AGN populations, we use a set of classical AGN diagnostics as reference. These include emission-line diagnostics (BPT-AGN), MIR colors (WISE-AGN), and high-energy diagnostics such as X-ray and radio emission. These methods are applied to galaxies at $z \leq 0.5$, where line measurements and MIR photometry are most reliable. 
While the MIR color-based AGN selection draws from a subset of the photometric data available to full SED fitting, it remains a valuable cross-check. Its inclusion in this analysis serves primarily as a benchmark, given its common use and well-calibrated thresholds (e.g., \citealt{Stern2012, Assef2013}). We do not position it on equal footing with spectroscopy-based diagnostics or full SED fitting, as the latter incorporate broader wavelength coverage and physical modeling. Nonetheless, MIR color selection remains relevant for assessing completeness and consistency in AGN selection across surveys. 
To construct the BPT diagrams and BL-AGN sample, we use the emission line measurements from the DESI EDR FastSpecFit Spectral Synthesis and Emission-Line Catalog ({\tt FastSpecFit} version 3.2 \textit{Fuji} production\footnote{\url{https://data.desi.lbl.gov/doc/releases/edr/vac/fastspecfit/}}; \citealt{Moustakas2023}; Moustakas et al. in prep.). 
 {\tt FastSpecFit} is a stellar continuum and emission-line fitting code optimized to model jointly DESI optical spectra and broadband photometry using physically motivated stellar continuum and emission-line templates\footnote{\url{https://fastspecfit.readthedocs.io/en/latest/fuji.html}}. The selection of the AGN sample is summarized in Sect.~\ref{sec:ParentSample}.

\subsection{BPT diagram}\label{sec:BPTdiagram}

 The BPT diagrams~\citep{Baldwin1981} are widely used diagnostics relying on the strength of Balmer lines to forbidden nebular transitions to differentiate between star-forming galaxies and AGN. 
 In this work, we use three diagrams based on the ratios of $\rm [NII]\lambda6583/H\alpha$ vs. $\rm [OIII]\lambda5007/H\beta$ ([NII]-BPT), $\rm [SII]\lambda6717,6731/H\alpha$ vs. $\rm [OIII]\lambda5007/H\beta$ ([SII]-BPT), and $\rm [OI]\lambda6583/H\alpha$ vs. $\rm [OIII]\lambda6300/H\beta$ ([OI]-BPT) similarly as~\cite{Mezcua2024Manga}. 
 The [NII]-BPT diagram is sensitive to metallicity, making AGN identification in metal-poor galaxies difficult~\citep[e.g.,][]{Groves2006,Cann2019}. The [SII]-BPT offers better separation at intermediate metallicities but struggles with composite galaxies. The [OI]-BPT is more robust across metallicities but relies on a weak line that's often hard to detect in faint or low-metallicity galaxies~\citep{Polimera2022}. 
 Recently, \citealt{Ji2020} proposed a simple reprojection of the [NII]-, [SII]- and [OI]-BPTs that shows the potential to remove the ambiguity for the true composite objects. We leave the examination of this approach for future work. 

 The [NII]-, [SII]-, and [OI]-BPTs for the DESI galaxies are shown in Fig.~\ref{fig:BPT}. 
 We use the narrow components of the Balmer lines and apply a flux SNR $\geq 3$ in each of the emission lines used for the respective [NII]-, [SII]-, and [OI]-BPT diagrams, except for $\rm [OI]\lambda6300$, for which we use a flux SNR $\geq 1$. The $\rm [OI]\lambda6300$ line is weaker, and imposing a higher SNR threshold could exclude valid candidates~\citep{CidFernandes2010}. The SNR cuts are applied separately for each diagram, meaning that the galaxies included in each BPT diagram may differ. 

We also use the WHAN diagram introduced by~\cite{CidFernandes2010} that considers equivalent width (EW) of $\rm H\alpha$ ($\rm EW(H\alpha)$) vs.  $\rm [NII]\lambda6583/H\alpha$ line ratios. The WHAN diagram differentiates AGN (with $\rm EW(H\alpha) \ge 3$ and $\rm log [NII]/H\alpha > -0.4$) from galaxies with hot old (post-AGB) stars that can mimic LINERs.  To ensure reliable measurements, we require a flux SNR $\geq 3$ for all relevant lines ($\rm H\alpha$ and $\rm [NII]\lambda6583$), as well as for $\rm EW(H\alpha)$. Consequently, the WHAN sample is not necessarily the same as the one used for the BPT diagrams, as the SNR cuts are applied independently.

\subsection{BL-AGN selection}\label{sec:BL_AGN_selection}

We identify BL-AGN at $z \leq0.5$  based on the presence of a broad H$\alpha$ component. Following the criteria outlined in~\cite{Pucha2024}, we require SNR $\geq 3$ for the total H$\alpha$ flux ($\tt SNR_{HALPHA\_FLUX}$), the broad H$\alpha$ component ($\tt SNR_{HALPHA\_BROAD}$), and the amplitude-over-noise (AoN) of the broad component ($\tt AoN_{HALPHA\_BROAD}$). %

\begin{figure*}
\centerline{\includegraphics[width=0.44\textwidth]{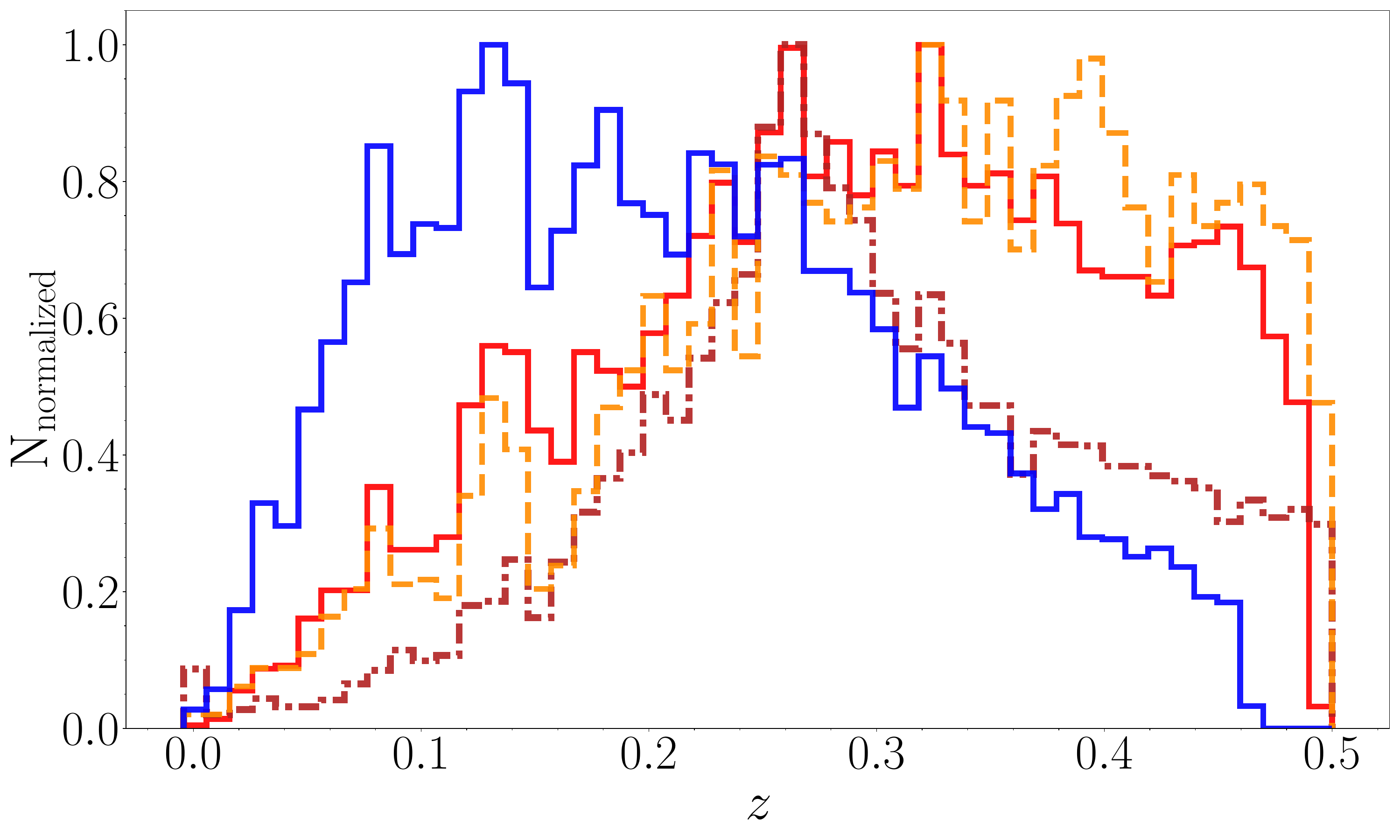}
\includegraphics[width=0.44\textwidth]{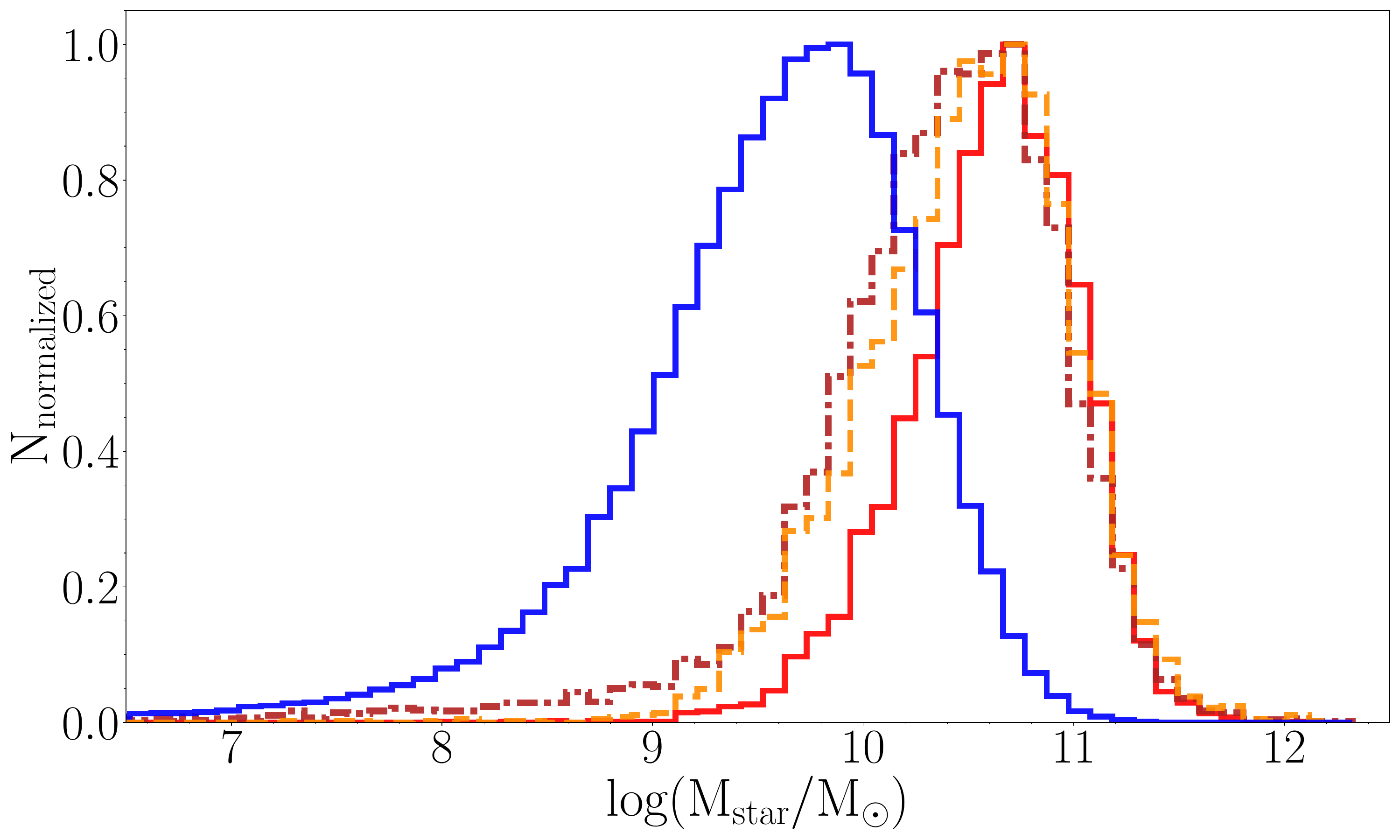}}
\centerline{\includegraphics[width=0.44\textwidth]{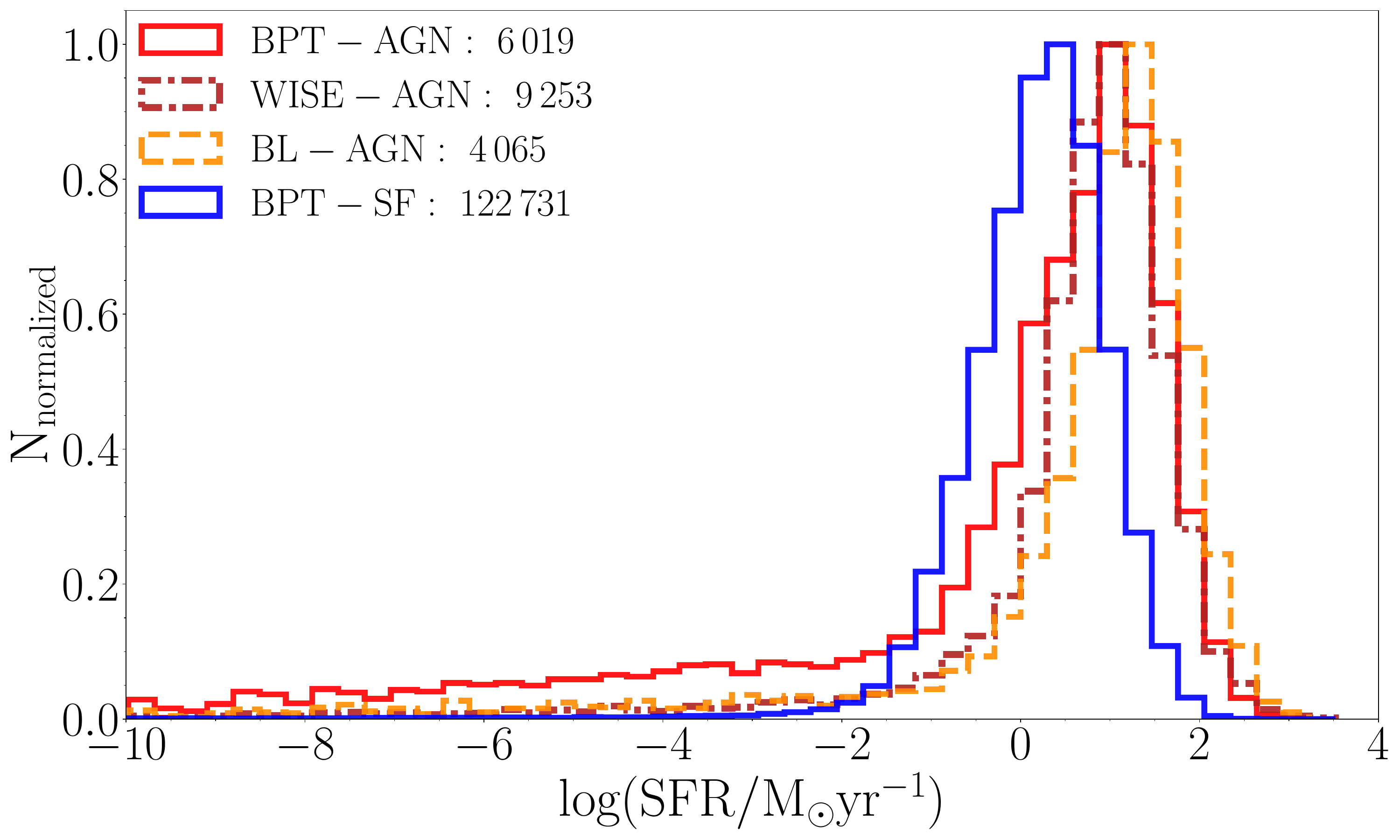}
\includegraphics[width=0.44\textwidth]{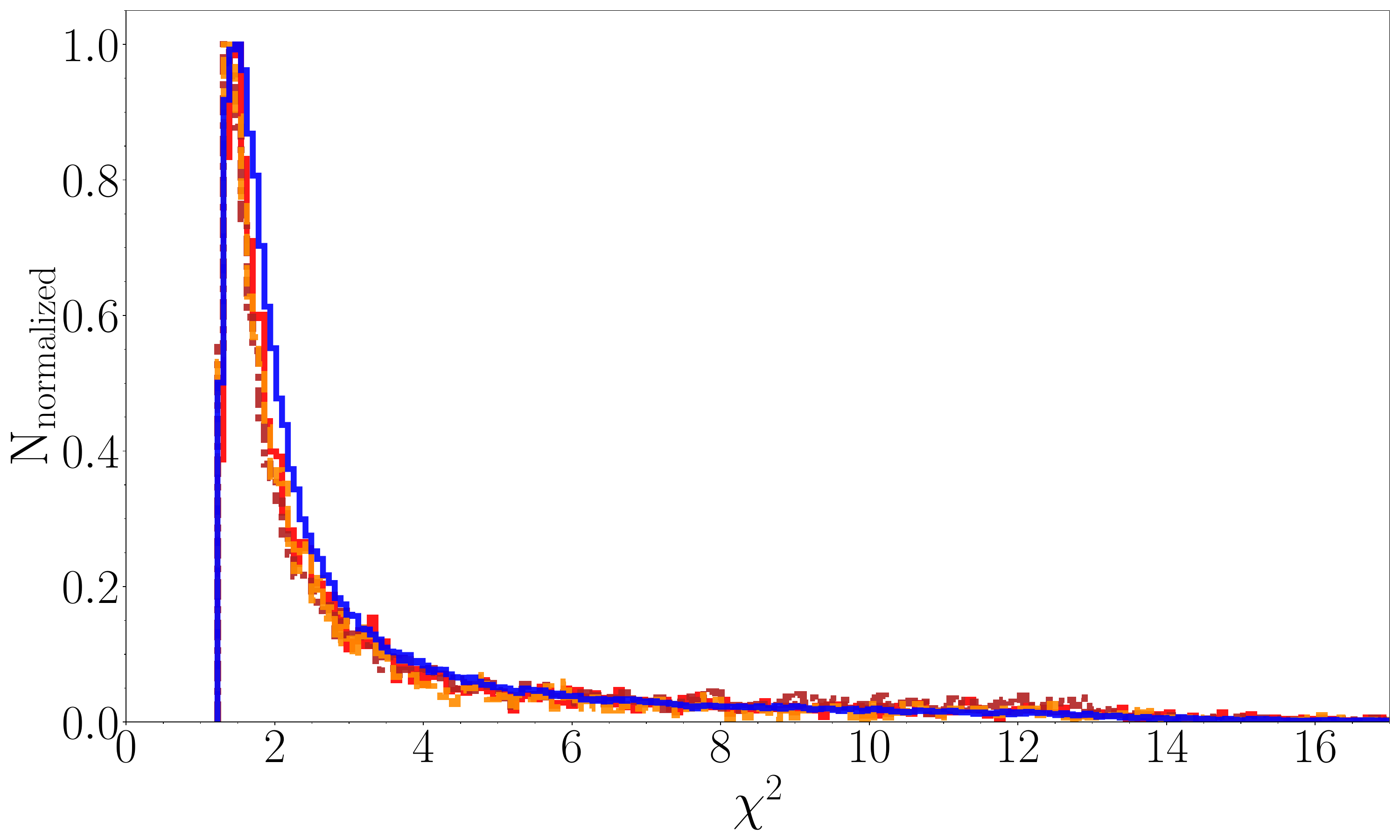}}
	\caption{Distribution of $z$ (top left panel), stellar mass (top right panel), SFR (bottom left panel), and $\chi^2$ (bottom right panel) for the AGN, and star-forming samples.}
	\label{fig:phys_prop_distrubution}
\end{figure*}  

 \subsection{WISE selection}\label{sec:WISEselection}

To select MIR-AGN we rely on the WISE color-color diagram $W2-W3$ vs. $W1-W2$ proposed by~\citep{Hviding2022}. We restrict the sample to galaxies with $\tt SNR \rm \ge 3$ in all WISE bands used in the diagram and to $z \le 0.5$.  We note, that the forced WISE photometry can suffer from confusion and other systematic errors, which can impact the quality of the fitting and the resulting inference as well as the WISE SNR cuts. In addition, Legacy Surveys WISE photometry is quite shallow, especially in $W3$, and $W4$ and especially for emission line galaxies and other higher-redshift galaxies, which impacts the ability to identify AGN from the broadband SED modeling. 
The selection criteria proposed by~\cite{Hviding2022} are more sensitive to low luminosity and heavily obscured AGN than the traditional cuts given by~\cite{Jarrett2011} or~\cite{Stern2012}. 
However, WISE-AGN samples may be contaminated by star-forming galaxies diluting the WISE color-selected AGN~\citep{Hainline2016A}.

\subsection{X-ray and radio counterparts}~\label{sec:xray_selection}

We search for X-ray counterparts within 5 arcsec of the optical position of each source as given by the DESI Legacy Imaging Surveys making use of the Chandra Source Catalog~\citep[CSC DR2;][]{Evans2010, Evans2024}, the XMM-Newton Serendipitous Source Catalogue~\citep[4XMM-DR13;][]{Webb2020}, and the all-sky survey catalog from the extended ROentgen Survey with an Imaging Telescope Array (eROSITA) on the Spectrum-Roentgen-Gamma (SRG) mission~\citep[eRASS1;][]{Merloni2024}. For identification of the radio counterparts within 5 arcsec from the sources we rely on the second data release of the LOFAR Two-metre Sky Survey~\citep[LoTSS DR2;][]{Shimwell2022}, the  Faint Images of the Radio Sky at Twenty-cm (FIRST) Survey Final Catalog~\citep{Helfand2015} and the 1.4 GHz National Radio Astronomy Observatory (NRAO) Very Large Array (VLA)  Sky Survey~\citep[NVSS;][]{Condon1998A}.

\subsection{AGN sample}\label{sec:ParentSample}

We apply the optical and MIR selection criteria described above to classify 510\,939 DESI EDR galaxies into four types. These four classes are used as a reference to estimate the best threshold on {\tt AGNFRAC} for identifying AGN based on the SED fitting: 
\begin{enumerate}
    \item BPT-AGN: 
    defined as AGN or composite in the [NII]-BPT,  AGN on the WHAN diagram, and AGN in the [SII]- or [OI]-BPT. 
    Selected number of sources: 6\,019.
    \item BPT-SF:
    defined as star-forming in the [NII]-, [SII]- and [OI]-BPTs. Selected number of sources: 122\,731.
    \item BL-AGN: 
    defined as sources with a broad $\rm H\alpha$ emission line. Selected number of sources: 4\,065.
    \item WISE-AGN: 
    defined as AGN based on the WISE color diagram. Selected number of sources: 9\,253.
\end{enumerate}

\begin{figure}
 	\centerline{\includegraphics[width=0.49\textwidth]{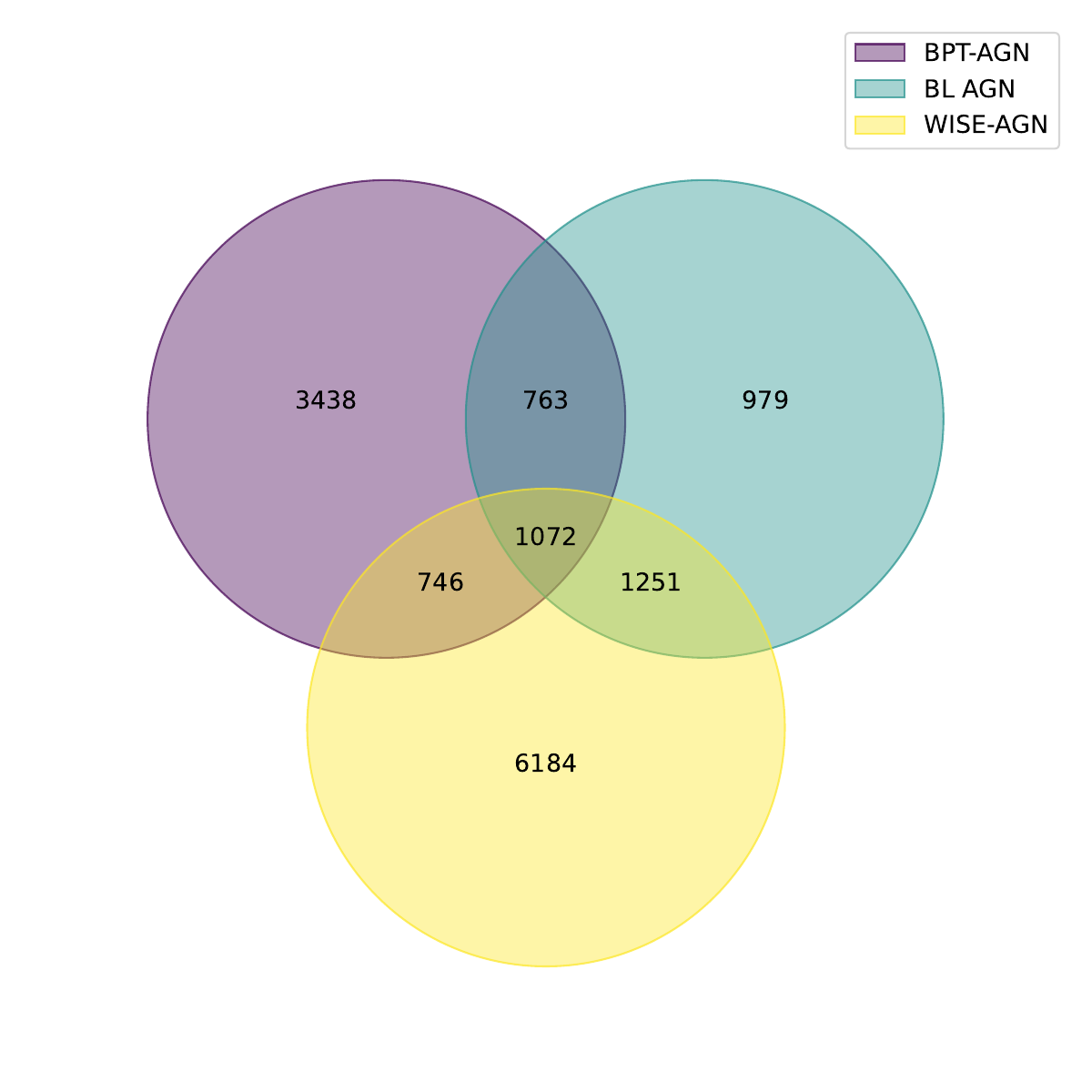}}
	\caption{Venn diagram illustrating the overlap between AGN selected using different methods: BPT-AGN (violet circle), BL-AGN (green circle), and WISE-AGN (yellow circle). The areas of overlap between circles indicate AGN identified by multiple methods, highlighting the diversity and overlap in AGN classification across different diagnostic techniques.}
	\label{fig:Venn}
\end{figure}

 Each class includes galaxies at $z\leq0.5$ spanning a wide range of stellar masses. As shown in Fig.~\ref{fig:phys_prop_distrubution}, the BPT-SF population predominantly consists of lower-mass, lower star-formation galaxies at lower redshifts. Additionally, since SFRs are derived from CIGALE using AGN-inclusive templates, even BPT-SF galaxies may show slightly suppressed SFRs if the fit assigns any AGN contribution -- this is especially relevant in ambiguous or composite systems. We explore also whether the $\chi^2$ distribution varies across different AGN and star-forming galaxy classifications (bottom right panel in Fig.~\ref{fig:phys_prop_distrubution}). 
Contrary to some earlier works~\citep[e.g.,][]{Berta2013} where elevated $\chi^2$ values -- especially in stellar or star-forming-only fits -- could signal the presence of a hidden AGN (e.g., visible in the NIR excess), we find that the distributions of reduced $\chi^2$ are remarkably similar for BPT-AGN, WISE-AGN, BL-AGN, and BPT-SF galaxies. This is likely due to the inclusion of AGN templates in the CIGALE fits, which improves the overall model match even in AGN-dominated sources. This suggests that $\chi^2$ is not a discriminant for AGN identification in our case, provided that proper AGN templates are used during the fitting process. 
Figure~\ref{fig:Venn} and Table~\ref{tab:agn_stats} highlight the diversity and overlap between AGN identified through different selection techniques: BPT-AGN, BL-AGN, and WISE-AGN.  The significant areas of non-overlap (e.g., 66.8\% of WISE-AGN are identified only by WISE colors) underscore that each method may be sensitive to different types of AGN and highlight the need for a multi-wavelength approach to identify the complete AGN sample as already investigated by e.g.,~\cite{Cann2019} and~\cite{Hviding2022}.

\begin{table*}[ht]
\centering
\begin{tabular}{lrrrrrrrrr}
\hline
{Classification} & $\rm N_{total}$ & $\rm N_{unique}$ & $\rm \%_{unique}$ & $\rm N_{BPT-AGN}$ & $\rm \%_{BPT-AGN}$ & $\rm N_{BL-AGN}$ & $\rm \%_{BL-AGN}$ & $\rm N_{BPT-SF}$ & $\rm \%_{BPT-SF}$ \\
\hline\hline
BPT-AGN & 6\,019 & 3\,438 & 57.1 & -- & -- & 1\,835 & 30.5 & 0 &  0.0 \\
\hline
BL-AGN & 4\,065 & 979 & 24.1 & 1\,835 & 45.1 & -- & -- & 103 & 2.5 \\
\hline
WISE-AGN & 9\,253 & 6\,184 & 66.8 & 1\,818 & 19.6 & 2\,323 & 25.1 & 1\,907 & 20.6 \\
\hline
\end{tabular}
\caption{Overlap between BPT-AGN, BL-AGN, WISE-AGN, and BPT-SF samples (see Sect.~\ref{sec:ParentSample}). The table summarizes the total number of identified AGN, the number and percentage of uniquely identified AGN, and overlap with other classification methods. }
\label{tab:agn_stats}
\end{table*}

\section{Results and Discussion}\label{sec:resultsanddiscussion}
In this Section, we validate the AGN fraction ({\tt AGNFRAC}) estimated via SED fitting of optical ({\it grz}) and MIR ({\it W14}) observations as a proxy to identify AGN in comparison to standard techniques. As a large fraction of AGN is not detected in optical surveys due to either distance or dust obscuration (e.g., \citealt{Truebenbach2017}), we validate how crucial it is to include the MIR photometry in SED fits to recreate the AGN contribution in galaxies. We also identify AGN candidates selected by the SED method that do not appear on traditional AGN diagrams.

\subsection{SED fitting: AGN fraction}\label{sec:AGNfraction}

 \begin{figure*}
 	\centerline{\includegraphics[width=0.99\textwidth]{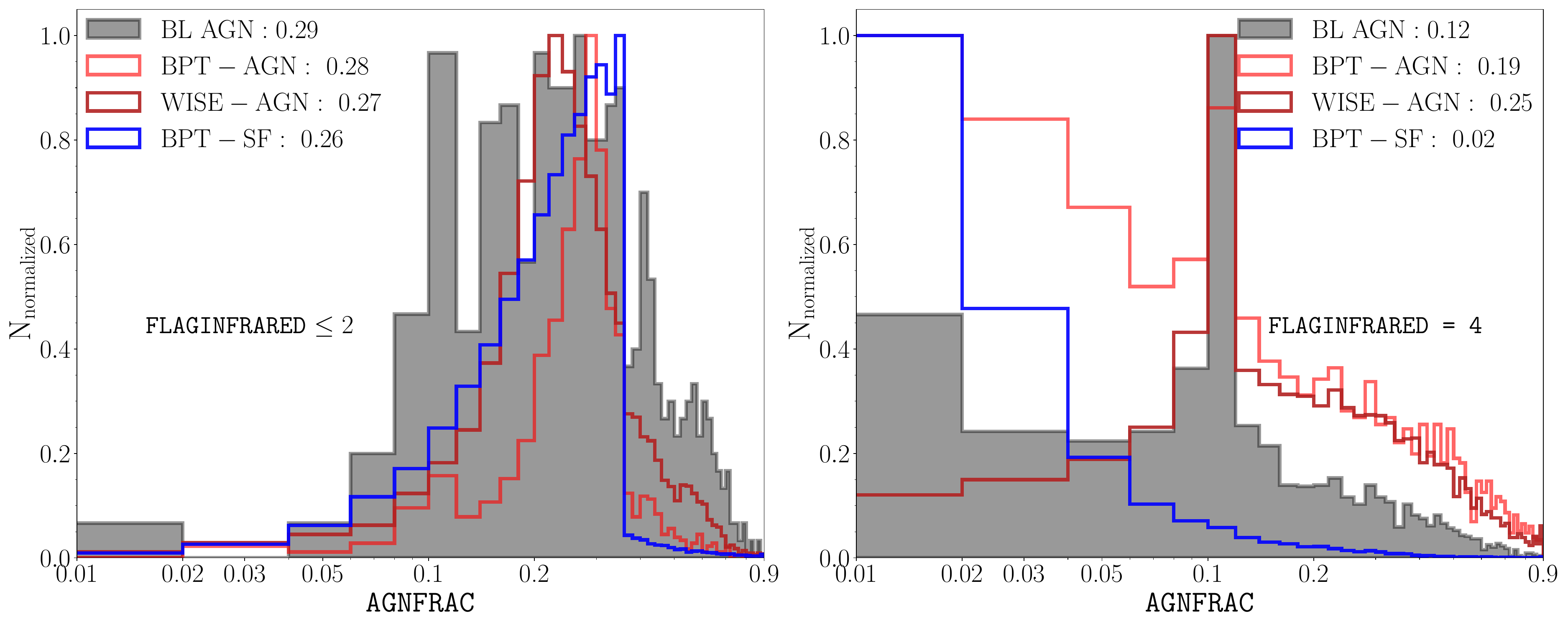}}
	\caption{Distribution of the AGN fraction defined as the fraction of the IR emission coming from the AGN to the total IR emission ({\tt AGNFRAC}) for BPT-selected star-forming galaxies (BPT-SF), BPT-selected AGN, WISE-selected AGN and BL-AGN observed in at most two WISE bands with $\tt SNR \ge 3$ ($\tt FLAGINFRARED \leq 2$; left plot) or at all four WISE bands ({\tt FLAGINFRARED} = 4; right plot).  The median {\tt AGNFRAC} are reported in the legend.  }
	\label{fig:AGNFrac}
\end{figure*}  

Usually, SED-based AGN selection is based on a fixed threshold on the AGN fraction derived from SED fitting~\citep{Thorne2022, Best2023,Bichang2024, Das2024}. The AGN fraction is defined here as the fraction of the IR emission coming from the AGN to the total IR emission. We rely on the AGN and star-forming galaxy sample selected based on the traditional techniques (see Sect.~\ref{sec:ParentSample}) to find the best threshold for the AGN selection. 
 Figure~\ref{fig:AGNFrac} shows the distribution of {\tt AGNFRAC} for galaxies with different infrared coverage. We split the sample based on the {\tt FLAGINFRARED} value, which indicates the number of WISE bands with high SNR ($\rm SNR \geq 3$): a value of 4 means detections in all four bands, while lower values correspond to fewer reliable measurements. Although {\tt AGNFRAC} uncertainties are computed by {\tt CIGALE}, they are not included in the DESI EDR VAC and in this analysis because, in cases with faint AGN emission or limited infrared data, the error estimates can become poorly constrained and unreliable for robust statistical use. This limitation is inherent to SED fitting in such regimes and does not affect the broader AGN selection performance (see also \citealt{Ciesla2015}, \citealt{Yang2020}, and \citealt{Mountrichas2021b}). 
In the left panel (galaxies observed at most in two WISE bands with $\rm SNR \geq 3$, i.e., with {\tt FLAGINFRARED} $\leq 2$), AGN and star-forming galaxies show nearly indistinguishable {\tt AGNFRAC} distributions (typical values around 0.3), indicating unreliable AGN fraction estimates when only one or two WISE bands are available. In contrast, the right panel (galaxies observed in all four WISE bands with $\rm SNR \geq 3$, i.e., with {\tt FLAGINFRARED} = 4) reveals clear separation: star-forming galaxies show low {\tt AGNFRAC} values (median = 0.02), while AGN classes (BPT-AGN, BL-AGN, WISE-AGN) have broader distributions with higher medians (0.19, 0.12, and 0.25, respectively). Notably, BPT-AGN exhibit a bimodal distribution peaking at 0 and ~0.1. 

 To further evaluate the discriminative power of {\tt AGNFRAC}, we perform a classification analysis using Receiver Operating Characteristic~\citep[ROC;][]{Bradley1997} curves across our samples. The performance metrics and threshold analysis (see Appendix~\ref{app:ROC}) suggest that {\tt AGNFRAC} is an effective AGN indicator only when applied to sources with reliable infrared measurements.

 Taking into account the strong correlation between {\tt AGNFRAC} and the availability of WISE photometry with ${\rm SNR} \geq 3$ (see also Sect.~\ref{sec:MIRphotometry}), we recommend applying a selection criterion of {\tt AGNFRAC} $\ge 0.1$ combined with {\tt FLAGINFRARED} $= 4$ to define AGN based on SED fitting (SED-AGN). This threshold is motivated both by the observed peak in {\tt AGNFRAC} distributions for spectroscopically confirmed AGN classes (see Fig.~\ref{fig:AGNFrac}) and by its optimal diagnostic performance in the ROC analysis (see Appendix~\ref{app:ROC}). For more inclusive samples, the infrared quality condition can be relaxed to {\tt FLAGINFRARED} $\geq 3$, which preserves the general distribution shape and yields similar median {\tt AGNFRAC} values.

\subsection{Importance of the MIR photometry}\label{sec:MIRphotometry}

\begin{figure*}
 	\centerline{\includegraphics[width=0.99\textwidth]{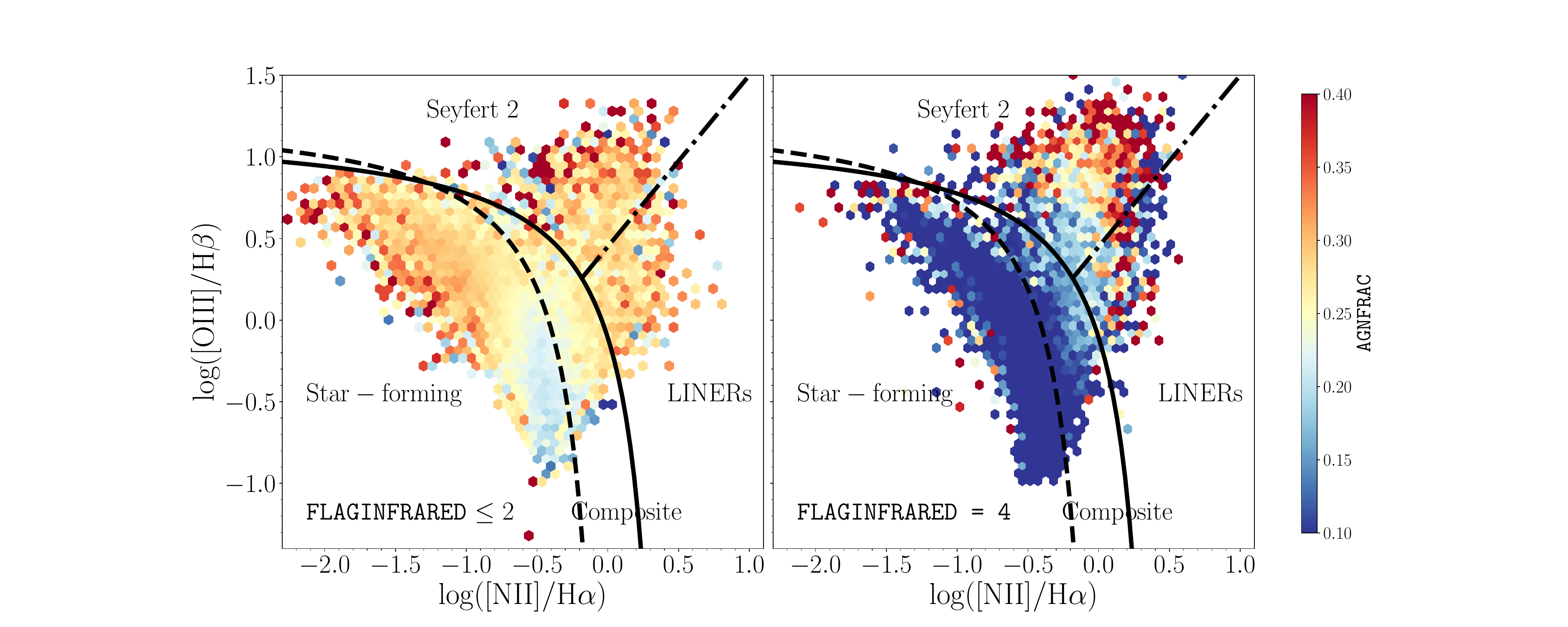}}
	\caption{BPT diagram for DESI galaxies observed at most in two MIR bands with SNR $\geq 3$ (i.e., {\tt FLAGINFRARED} $\le 2$; left panel) and with all four WISE photometry with SNR $\geq$ 3 (i.e., {\tt FLAGINFRARED} = 4; right panel). The demarcation lines separating star-forming galaxies, composites, AGN, and LINERs as proposed by \cite{Kewley2001,Kauffmann2003} and \cite{Schawinski2007} are marked with dashed, solid, and dash-dotted lines, respectively. The plots are color-coded by the AGN fraction ({\tt AGNFRAC} is defined as the fraction of the IR coming from the AGN to the total IR emission). The AGN fraction increases while moving to the BPT region associated with AGN and LINERs when all four WISE bands with  $\rm SNR \geq 3$ are provided in the SED fits. Lack of  $\rm SNR \geq 3$ in the four MIR bands may result in an artificial overestimation of the AGN fraction ({\tt AGNFRAC} $\geq 0.1$) for a majority (62\%) of star-forming galaxies. }
	\label{fig:BPT_AGNfraction}
\end{figure*}  

As shown in the previous Section, the distribution of {\tt AGNFRAC} is sensitive to the availability of WISE photometry. We further verify the dependence of the BPT classification on the availability of the WISE photometry. Figure~\ref{fig:BPT_AGNfraction} shows the [NII]-BPT diagram for galaxies observed in all four WISE bands with $\tt SNR \rm \ge 3$ (i.e, {\tt FLAGINFRARED} = 4; right panel) and galaxies with at most two WISE observations with $\tt SNR \rm \ge 3$ (i.e, {\tt FLAGINFRARED} $ \le 2$; left panel). 
The distribution of {\tt AGNFRAC} is almost uniform on the BPT diagram for galaxies without high SNR WISE photometry. 
Contrarily, for galaxies with {\tt FLAGINFRARED} = 4, the {\tt AGNFRAC} is gradually changing from $\lesssim 0.1$ in the star-forming galaxies locus to $\gtrsim 0.4$ in the region attributed to AGN. 

{Using the AGN fraction derived with {\tt CIGALE} as the criterion to select AGN ({\tt AGNFRAC} $\geq0.1$), we recover 74\% of BPT-selected AGN (69\% if limiting to sample with {\tt FLAGINFRARED} = 4). However, at the same time, 62\% star-forming galaxies also are identified as AGN (i.e., fulfilling the criterion {\tt AGNFRAC} $\geq0.1$). The percentage of misclassified star-forming galaxies drops to 15\% for a sample with {\tt FLAGINFRARED} = 4.}
We note that~\cite{Siudek2024} show that the overestimation of the AGN fraction for misclassified star-forming galaxies (with {\tt AGNFRAC}  $\geq 0.1$) does not affect the estimates of the stellar masses and SFRs (see Sect.~\ref{sec:DesiEDR}).

\begin{figure*}
 	\centerline{\includegraphics[width=0.99\textwidth]{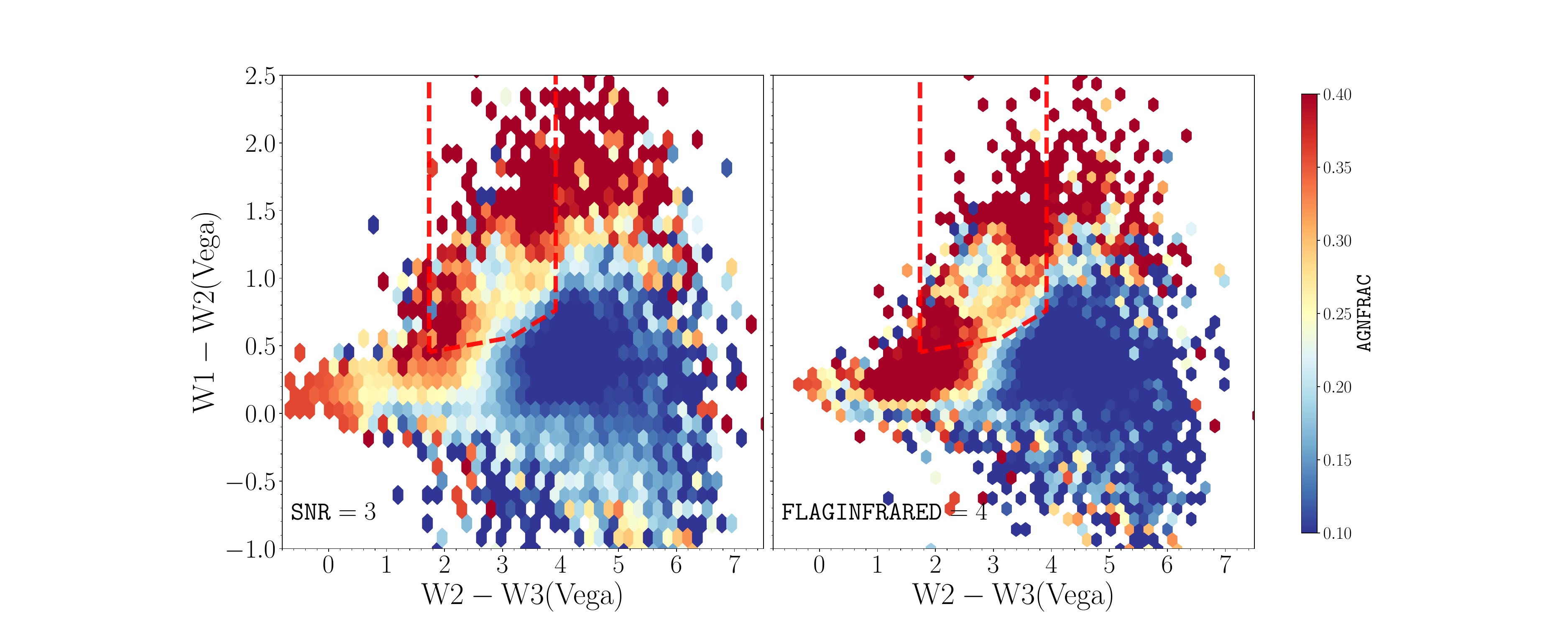}}
	\caption{WISE diagram as a function of AGN fraction for DESI galaxies observed with $\rm SNR \ge 3$ at $W1-3$ (left panel) or at all WISE bands (i.e., {\tt FLAGINFRARED} = 4; right panel). The region of AGN is marked by a red dotted line and follows the criterion proposed by \cite{Hviding2022}. The WISE-selected AGN are characterized by higher AGN fraction than non-AGN galaxies except for the SED-MIR-AGN candidates, which are characterized by a high AGN fraction ({\tt AGNFRAC} $\gtrsim 0.5$ at the right panel). We note that 96\% of these SED-MIR-AGN missed by the WISE diagram are identified as AGN by the TBT diagram. }
	\label{fig:WISE_AGNfraction}
\end{figure*}

Figure~\ref{fig:WISE_AGNfraction} shows the WISE diagram as a function of the {\tt AGNFRAC} depending on the availability of the WISE photometry. There is no visual difference in the dependence on the {\tt AGNFRAC} whether the sample is limited to $\rm SNR \ge 3$ in {\it W1-3}; i.e., {\tt FLAGINFRARED} = 3 or all WISE bands ({\it W1-4}; i.e., {\tt FLAGINFRARED} = 4), but the sample size differs significantly.  The sample selected with {\tt FLAGINFRARED} $\geq 3$ includes 9\,253 sources, while restricting to {\tt FLAGINFRARED} = 4 results in 4\,974 galaxies. 
{Without any restriction on the {\tt FLAGINFRARED}, the WISE-AGN are recovered well with SED selection (79\% of WISE-AGN are identified as SED-AGN). At the same time, 40\% of non-AGN are identified as SED-AGN. Only when applying a more strict cut on WISE photometry ({\tt FLAGINFRARED} = 4) leads to the identification of 86\% of WISE-AGN and reduces the misclassified AGN to 33\%. }

We find a similar correlation of the {\tt AGNFRAC} with WISE colors as found for the BPT selection: the region attributed to AGN  is characterized by a high AGN fraction (with a median {\tt AGNFRAC} = 0.25). 
Outside the AGN selection region, the AGN fraction is low ({\tt AGNFRAC} $\lesssim 0.07$). However, there are two regions attributed to higher AGN fraction ({\tt AGNFRAC} $\sim 0.5$) on the right-low end and left-upper of the AGN envelope. While the upper-left region is sparsely populated and can be neglected, the tail in {\it W2-W3} ($W2-W3 \lesssim 2.5$) across $ 0.5 \lesssim W1-W2 \lesssim 0$ color gathers 2\,017 galaxies (i.e., half of the WISE-AGN sample; SED-MIR-AGN candidates hereafter). 

SED-MIR-AGN candidates are in the majority (91\%) characterized by high AGN fraction ({\tt AGNFRAC} $\gtrsim 0.5$)\footnote{We consider the sample restricted to {\tt FLAGNFRARED} =4.}. 
They are massive passive galaxies (with median $\mstar \sim 11$ and median $\rm log(SFR/M_{\odot}y^{-1}) \sim -6$). 
SED-MIR-AGN candidates show tentative\footnote{As $\neIIIlam$ is weak and noisy we do not restrict the sample based on their SNR.} signatures of $\neIIIlam$ emission (with median full width at half maximum similar to the one found for AGN, i.e., FWHM $\rm \sim 320$ km~$\rm s^{-1}$, while for non-AGN FWHM is $\rm \sim 230$ km~$\rm s^{-1}$). 
The [NeIII]$\lambda$3869 emission together with $\oiilam$ emission line and {\it g-z} rest-framed color forms another AGN diagnostic diagram (TBT diagram; \citealt{Trouille2011}). \cite{Trouille2011} showed that the TBT diagram is efficient in recreating [NII]-BPT selection (98.7\% of BPT-AGN are identified as TBT-AGN and 97\% of the BPT-SF as TBT-SF) at the same time outperforming the BPT in identifying X-ray selected AGN (TBT identifies 97\% of the X-ray AGN as TBT-AGN, while BPT only 80\%). Almost all SED-MIR-AGN candidates (94\%) are identified as AGN according to the TBT diagram, however, we cannot be confident of their AGN nature, as only 65 have $\neIIIlam$ detected with $\rm SNR\geq 3$. The weak and noisy $\neIIIlam$ may bias the conclusion.

\subsection{SED-AGN sample}~\label{sec:SED-AGNsampleComparison}

Table~\ref{tab:sed_agn_efficiency_parent_sample} shows that the SED-AGN sample, defined by galaxies with {\tt AGNFRAC}~$\geq0.1$ and {\tt FLAGINFRARED}$=4$ at $\rm z\leq 0.5$, identifies a significant number of AGN ($\gtrsim 70\%$) selected using traditional methods. This overlap suggests that the SED-fitting approach is capable of capturing a broad range of AGN types. As detailed in Table~\ref{tab:sed_agn_efficiency_parent_sample}, the SED-based classification successfully identifies around 70\% of BL-AGN and BPT-AGN and over 85\% of WISE-AGN. This high level of correspondence demonstrates the robustness of the SED method in capturing classical AGN types. 
However, about 15\% of star-forming galaxies (BPT-SF) are also classified as AGN by the SED method, indicating some level of contamination from non-AGN sources.

A particularly notable result is that over half of the SED-AGN sample ($\sim52\%$) is not identified by any of the main AGN selection methods (BL-AGN, BPT-AGN, WISE-AGN, or BPT-SF). This raises an important question: is the SED-based method producing a high rate of false positives, or is it effectively uncovering a population of AGN that are missed by traditional diagnostics? 
To better understand the nature of the SED-AGN sample, we expanded our analysis beyond the standard AGN selection techniques. Specifically, we incorporate additional AGN classifications derived from all three BPTs, X-ray, and radio AGN diagnostics. This broader approach allows for a more comprehensive assessment of the AGN content within the SED-selected population. 

When incorporating these additional AGN identification methods, the fraction of galaxies classified solely by the SED criteria (SED-AGN Only) decreases significantly -- from 51.78\% to 16.14\% (see Table~\ref{tab:sed_agn_efficiency_detailed}), highlighting both the complementary value and the limitations of the SED-based method. The SED-based approach appears to at least partially recover AGN populations that are overlooked by traditional diagnostics, emphasizing the value of SED fitting in building a more complete and inclusive AGN census. However, at the same time, a non-negligible fraction of the SED-AGN sample ($\sim15\%$) overlaps with star-forming galaxies (BPT-SF), suggesting some level of contamination. Moreover, the AGN nature of sources classified solely as SED-AGN remains unconfirmed (see discussion below), challenging the use of SED fitting for constructing clean and pure AGN samples. 

However, we note that the nature of LINERs remains debated, with alternative ionization sources such as post-asymptotic giant branch (post-AGB) stars or shocks being proposed in the literature (e.g. Stasińska et al. 2008; Singh et al. 2013; Belfiore et al. 2016).

 The SED-based technique identifies roughly 50\% of LINERs as selected by the BPT diagrams. The interpretation of LINERs remains actively debated in the literature, with studies supporting both AGN~\citep[e.g.,][]{Heckman1980, Kewley2006} and stellar photoionization scenarios~\citep[e.g.,][]{Stasinska2006, Sarzi2010}. The overlap with LINERs suggests that the SED-AGN method is sensitive to a diverse AGN population, including those in different evolutionary stages or in obscured environments. The SED-based method maintains a relatively low level of contamination except in the WHAN diagram. 
The SED-AGN identifies 40\% of WHAN-AGN but also 67\% of retired galaxies. 
Interestingly, the majority (96\%) of these retired SED-AGN are actually "liny" (i.e., showing the presence of the emission lines; $\rm 0.5 \leq (EW(H\alpha) <3~\AA$) galaxies.  
The WHAN diagram is known to struggle with distinguishing low-luminosity AGN from "liny" retired galaxies, where line emission could originate from past AGN activity or low-level star formation~\citep{CidFernandes2010, CidFernandes2011, Herpich2018}. 
This raises the possibility that the SED-AGN method may be identifying galaxies transitioning from AGN activity to a more passive state, thus appearing as retired on the WHAN diagram. 
Supporting this interpretation, \cite{Herpich2018} suggested that "liny" retired galaxies might have undergone recent star formation, and their emission line gas could originate from the AGN activity. Similarly, \cite{Agostino2023M} show that nearly one-third of X-ray AGN fall in the WHAN "retired" locus. In our sample, out of 2\,386 SED-AGN classified as retired in WHAN, 515 (22\%) are independently confirmed as AGN via other diagnostics, further questioning their classification as fully passive systems.

The X-ray data (see Sect.~\ref{sec:xray_selection}) reveals 882 sources with X-ray emission. Using {the star formation rates derived from {\tt CIGALE} SED fitting and} the relation from~\cite{Lehmer2010}, we estimate the expected 2–10 keV luminosity from star formation and find that 793 galaxies exceed this by more than $3\sigma$, confirming AGN-like X-ray emission~\citep[see also][]{Mezcua2018}. The SED-based method identifies $\sim50\%$ of these X-ray AGN, even though X-ray data are not used in the fitting, demonstrating good sensitivity to AGN across different regimes. To understand why the other 50\% are missed, we examine their classification using other methods. Most are also missed by traditional BPT, and WISE diagnostics, with nearly 40\% falling into the WHAN-AGN region. This suggests that the missed X-ray AGN are generally low-luminosity or obscured, escaping both SED and classical diagnostics, and reinforces the known difficulty of identifying such AGN without deep X-ray data. This is consistent with previous studies such as \citet{Trouille2011} and \citet{Juneau2011, Juneau2013}, which showed that only $\sim20$–$30\%$ of X-ray AGN fall in the AGN region of the BPT diagram, with many appearing as star-forming or composite systems instead. These studies concluded that emission-line diagnostics become increasingly incomplete for identifying obscured or radiatively inefficient AGN. Our results are in agreement with these findings, and show that SED fitting adds significant value by recovering $\sim50\%$ of X-ray AGN,  highlighting its effectiveness in such regimes. 

\begin{table}
\centering
\begin{tabular}{lccc}
\hline
Category & N & SED-AGN & SED-AGN \\
& & Matches \% & Only \%  \\
\hline\hline
{\tt BL-AGN} & 2\,253 & 70.13  \\
{\tt BPT-AGN} & 2\,703 & 69.29  \\
{\tt WISE-AGN} & 4\,953 & 86.45 \\
{\tt BPT-SF} & 20\,717 & 15.25 \\
{\tt SED-AGN} & 25\,123 & - & 51.78 \\
\hline
\end{tabular}
\caption{Efficiency of the SED-AGN classification relative to the standard AGN identification methods, where N gives the number of AGN identified by a given method among galaxies with {\tt FLAGINFRARED} $=4$ at $\rm z\leq 0.5$. SED-AGN Matches (\%) indicates the percentage of galaxies within each category also identified as SED-AGN. SED-AGN Only (\%) refers to the percentage of objects that are classified as AGN solely by the SED criterion.   }
\label{tab:sed_agn_efficiency_parent_sample}
\end{table}

Of 13\,450 galaxies with radio detections (see Sect.~\ref{sec:xray_selection}), 5\,326 exhibit radio luminosities either above $\rm 5\times10^{23} W/Hz$ or more than 3$\sigma$ above what is expected from star formation (e.g., \citealt{Mezcua2019}), classifying them as radio AGN.  The SED fitting recovers 40\% of these sources. Similar to the X-ray case, the radio AGN missed by SED tend to be weakly active or obscured, with most showing no overlap with BPT, or WISE diagnostics, and  $\sim45\%$ falling within the WHAN-AGN region. This again highlights the complementary but incomplete nature of the SED-based classification and reinforces the need for multi-wavelength data to uncover the full AGN population. 
Moreover, the SED fitting depends on model assumptions, which affect the AGN classification (see Table~\ref{tab:model_dependence} in Appendix~\ref{app:model_dependence} for details).

\begin{table}
\centering
\begin{tabular}{lccc}
\hline
Category & N & SED-AGN & SED-AGN \\
& & Matches \% & Only \%  \\
\hline\hline
{\tt NII-AGN} & 3\,791 & 67.95 \\
{\tt NII-SF} & 23\,175 & 17.48  \\
{\tt NII-LINER} & 1\,064 & 54.42 \\
{\tt NII-COMPOSITE} & 7\,939 & 33.54 \\
\hline
{\tt SII-AGN} & 2\,583 & 68.45 \\
{\tt SII-SF} & 31\,458 & 21.13  \\
{\tt SII-LINER} & 630 & 53.65 \\
\hline
{\tt OI-AGN} & 3\,510 & 69.66 \\
{\tt OI-SF} & 30\,420 & 22.72  \\
{\tt OI-LINER} & 832 & 62.26 \\
\hline
{\tt WHAN-AGN} & 32\,283 & 38.61 \\
{\tt WHAN-SF} & 25\,534 & 20.36  \\
{\tt WHAN-RG} & 3\,559 & 67.04 \\
\hline
{\tt X-RAY-AGN} & 793 & 45.27 \\
{\tt RADIO-AGN} & 5\,326 & 38.64 \\
{\tt SED-AGN} & 4\,054 & - & 16.14 \\
\hline
\end{tabular}
\caption{Expanded evaluation of the SED-AGN classification (see Tab.~\ref{tab:sed_agn_efficiency_parent_sample}). The SED-AGN Only is significantly reduced when a more detailed AGN classification is applied including AGN ({\tt AGN}), star-forming galaxies ({\tt SF}), LINERs ({\tt LINER}), composites objects ({\tt COMPOSITE}), and retired galaxies ({\tt RG}) identified among [NII]-, [SII]-, [OI]- BPT, WHAN, radio and X-ray diagnostics.  }
\label{tab:sed_agn_efficiency_detailed}
\end{table}

\begin{figure*}
 	\centerline{\includegraphics[width=0.99\textwidth]{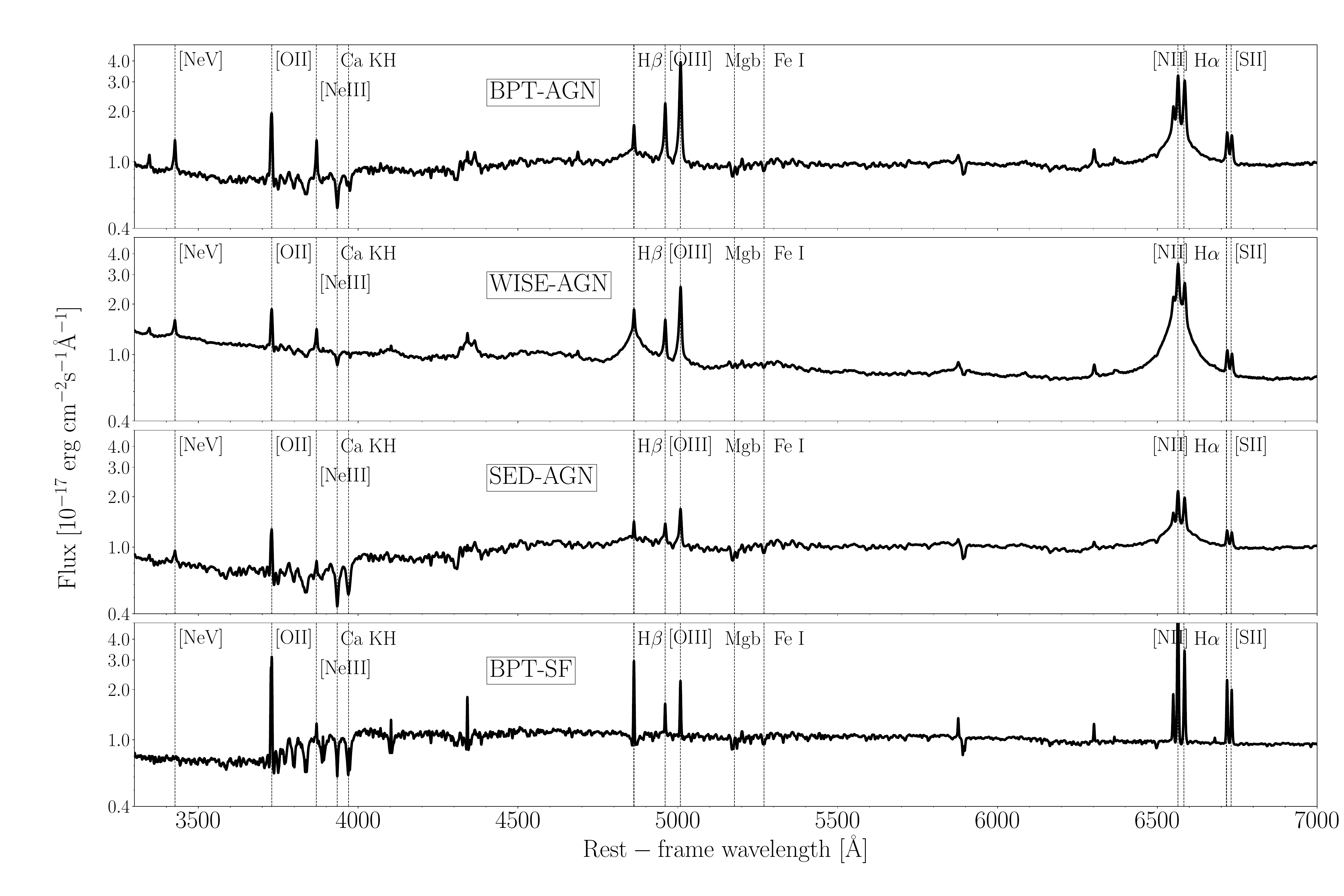}}
	\caption{Stacked spectra of BPT-AGN, WISE-AGN, SED-AGN, and BPT-SF. The stacked spectrum of SED-AGN galaxies highlights the characteristic AGN emission line ratios, suggesting their AGN nature. The stacked spectrum is created using the mean for normalization, preserving the relative fluxes of the emission lines. }
	\label{fig:spectrum}
\end{figure*}

Figure~\ref{fig:spectrum} shows the stacked spectra of BPT-AGN, WISE-AGN, SED-AGN, and BPT-SF galaxies. The spectra were stacked using mean normalization, preserving the relative fluxes of the emission lines, which is essential for a reliable comparison of spectral features across different samples. 
The SED-AGN stack shows similarity to the BPT-AGN and WISE-AGN stacks, particularly evident in the characteristic AGN emission line ratios, such as $\rm [NII]/H\alpha$. 
Unlike the BPT-SF sample, where the $\nevlam$ emission is absent, the presence and relative strength of $\nevlam$ in SED-AGN align more closely with the AGN-dominated spectra. 
High-ionization lines, such as $\nevlam$,  require energies well above the limit of stellar emission~\citep[55 eV; e.g.,][]{Leitherer1999,Izotov2004} and can be an efficient tracer of the AGN~\citep[e.g][]{Gilli2010, Negus2023, Barchiesi2024}.

\subsection{SED-AGN Only: properties of the host galaxies}\label{sec:PropHost}

Despite the effectiveness of the SED-based technique, 4\,054 out of 25\,123 SED-AGN galaxies (16.14\%, see Table~\ref{tab:sed_agn_efficiency_detailed}) are classified as SED-AGN Only, meaning they are not confirmed by any other AGN selection method. To further explore the nature of these SED-AGN Only galaxies, we analyze their properties in comparison to properties of the BPT-AGN, WISE-AGN, SED-AGN, and BPT-SF galaxies\footnote{We do not include BL-AGN in the comparison as the SED-AGN Only do not show the presence of broad emission lines.} as summarized in Table~\ref{tab:comparison_properties}. These SED-AGN Only galaxies tend to be more massive, with median stellar masses comparable to those of BPT-AGN and WISE-AGN hosts ($\rm log(M_{star}/M_{\odot}) \sim 10.75$). However, they exhibit significantly lower star formation rates ($\rm log(SFR/M_{\odot}y^{-1}) \sim -2$) than BPT-AGN and WISE-AGN, suggesting that they are generally passive galaxies.

To confirm the passive nature of SED-AGN Only hosts, we also check the 4000~\AA\ break (D4000),  an indicator of the age of the stellar population: a high D4000 value (D4000 > 1.5) indicates older stars and high metallicity, while a lower value (D4000 $\leq$ 1.5) suggests younger, star-forming populations~\citep[e.g.,][]{Kauffmann2003,Siudek2017, Siudek2018}. The high D4000 values of the SED-AGN Only (D4000 = 1.76) indicate that old stars dominate the light from these galaxies. 
Moreover, we find that the majority (78\%) of the AGN-SED Only show red colors on to the {\it U-V} vs. {\it V-J} diagram~\citep[][]{Whitaker2012, Siudek2024}, consistent with their passive nature.  This contrasts sharply with other classes, where the fraction of red galaxies is much lower, ranging from 0\% in star-forming galaxies to 13\% in BPT-AGN. 
Also, the optical part of the SEDs of SED-AGN Only galaxies is dominated by old stellar populations, while AGN contribution is seen in the MIR (see Appendix~\ref{app:SEDExemplary}).
 This analysis suggests that SED-AGN Only galaxies may represent a population of galaxies transitioning away from active star formation, possibly hosting weak or fading AGN activity that other methods fail to detect.

\begin{table*}[h!]
\centering
\begin{tabular}{lcccccc}
\hline
 & Unit & BPT-SF & BPT-AGN & WISE-AGN & SED-AGN & SED-AGN \\
 &  &  &   & & & Only \\
\hline
{\tt N} &  & 20\,717 &  2\,703 &  4\,953 & 25\,123 & 4\,054 \\
{\tt  Z} & & $0.19^{+0.09}_{-0.08}$ & $0.28^{+0.09}_{-0.09}$ & $0.29^{+0.09}_{-0.06}$ & $0.31^{+0.09}_{-0.08}$ & $0.29^{+0.12}_{-0.09}$ \\
{\tt LOGM} & $\rm [M_{\odot}]$ & $9.82^{+0.33}_{-0.37}$ & $10.61^{+0.25}_{-0.28}$ & $10.48^{+0.32}_{-0.38}$ & $10.58^{+0.25}_{-0.29}$ & $10.75^{+0.26}_{-0.31}$ \\
{\tt LOGSFR} & $\rm [M_{\odot} yr^{-1}]$ & $0.87^{+0.37}_{-0.40}$ & $1.10^{+0.42}_{-0.53}$ & $1.14^{+0.41}_{-0.46}$ & $0.69^{+0.52}_{-1.31}$ & $-2.23^{+2.10}_{-3.33}$ \\
{\tt D4000} & & $1.22^{+0.06}_{-0.05}$ & $1.33^{+0.17}_{-0.20}$ & $1.15^{+0.14}_{-0.12}$ & $1.22^{+0.14}_{-0.11}$ & $1.76^{+0.23}_{-0.32}$ \\
\hline
\end{tabular}
\caption{Comparison of the main properties: redshift, stellar masses ({\tt LOGM}) and SFRs ({\tt LOGSFR}) from the VAC of physical properties, and D4000 from {\tt FastSpecFit}. The median values and the errors (difference between the median and the 1st and 3rd quartiles) are given as well as the number of galaxies ({\tt N}) in each sample. 
}
\label{tab:comparison_properties}
\end{table*}

We also verify whether there are signatures of non-AGN nature among SED-AGN Only. Only 15 of SED-AGN Only ($< 1\%$) are characterized by line ratios characteristic of shock excitation, such as $\rm [SII](\lambda6717 + \lambda6731)/H\alpha > 0.4$ or $\rm [OI]/H\alpha>0.1$, which are traditionally associated with supernova remnants or other shock-heated regions rather than photoionization by an AGN~\citep[e.g.,][]{Dodorico1978, Rich2010, Comerford2022}. 
This suggests that contamination from non-AGN processes in the SED-AGN Only sample is minimal.

\section{Summary}\label{sec:conclusions}
We identify AGN based on the AGN fraction ({\tt AGNFRAC}~$\geq0.1$) derived from SED fitting with {\tt CIGALE} applied to DESI EDR galaxies at $z \leq 0.5$. As discussed in Sect.~\ref{sec:introduction}, traditional AGN identification diagnostics, such as the BPT diagram, are limited by luminosity biases, often leading to incomplete samples.  In contrast, the SED-based approach, leveraging a multi-wavelength strategy, promises a more comprehensive identification of AGN. Our key findings are summarized as follows:

\begin{itemize}
    \item Importance of WISE photometry:\\
    MIR photometry is essential for robust AGN identification via {\tt AGNFRAC}. When high SNR ($\rm SNR \geq 3$) WISE photometry is limited (i.e., {\tt FLAGINFRARED} $\leq 2$), 62\% of BPT-selected star-forming galaxies are incorrectly classified as AGN. Including all four WISE bands with $\rm SNR \geq 3$ ({\tt FLAGINFRARED} = 4) reduces this contamination to 15\%. Meanwhile, 70\% of BPT-AGN are consistently recovered, irrespective of the WISE photometry's SNR (see Sect.~\ref{sec:MIRphotometry}). 
    \item MIR-AGN candidates beyond WISE diagram::\\
    The SED fitting method identifies a distinct subset of strong AGN candidates (SED-MIR-AGN) that fall outside the canonical AGN region on the WISE diagram. These sources are confirmed as AGN in 98\% of cases via the TBT diagram, although weak and noisy $\neIIIlam$ emission limits definitive confirmation (see Sect.~\ref{sec:MIRphotometry}). 
    \item Comparison with standard AGN selection methods:\\   
    Our SED-based AGN classification shows substantial agreement with traditional techniques, recovering approximately 70\% of both BL-AGN and BPT-AGN, and over 85\% of WISE-selected AGN. However, the method does have a 15\% contamination rate from star-forming galaxies. 
    \item SED fitting as a unifying AGN diagnostic:\\
    The SED approach uncovers a population of AGN (SED-AGN Only) missed by standard diagnostics.   Relying solely on BPT-AGN, WISE-AGN, and BPT-SF selections fails to identify $\sim$52\% of SED-AGN (see Tab.~\ref{tab:sed_agn_efficiency_parent_sample}). Incorporating multiple selection methods reduces this fraction to $\sim$16\% (see Table~\ref{tab:sed_agn_efficiency_detailed}). While promising, the SED-AGN Only population includes uncertain cases (e.g., LINERs, “liny” retired galaxies), and may still be affected by a $\sim$15\% contamination from star-forming galaxies.
    \item Dependence on model assumptions:\\
    The efficiency and purity of SED-based AGN selection are moderately sensitive to modeling choices (see Table~\ref{tab:model_dependence} and Fig.~\ref{fig:ROC}). Changes to dust attenuation laws and AGN templates can alter the balance between completeness and contamination, while variations in IMF, SFH, and SSP models have smaller effects. We note that allowing metallicity to vary as a free parameter achieves similar or even slightly better performance, further reducing contamination from star-forming galaxies and marginally improving AGN selection efficiency. The use of mid-infrared data is crucial—removing WISE W3 and W4 bands leads to severe contamination by star-forming galaxies.
\end{itemize}

 This work demonstrates the importance of multi-wavelength data fusion for robust AGN identification, with direct implications for upcoming large-scale surveys. The Euclid mission will provide deep near-infrared photometry across 14\,000 $\rm deg^2$~\citep{Mellier2025}, LSST will deliver unprecedented optical depth and time-domain coverage~\citep{Ivezic2019}, and the Spectro-Photometer for the History of the Universe, Epoch of Reionization, and ices Explorer (SPHEReX) will perform an all-sky spectroscopic survey with near-immediate public data release~\citep{Dore2014, Crill2020}. However, mismatched depths and sky coverage among these facilities present challenges for optimal data fusion~\citep[e.g.,][]{Melchior2021, Huertas2023}.

Our 70-80\% recovery rate using DESI Legacy Surveys and WISE photometry demonstrates that systematic multi-wavelength approaches can effectively overcome individual survey limitations. To further enhance AGN selection, machine-learning techniques are proving  to be increasingly powerful. Semi-supervised clustering applied to medium-resolution spectroscopy can automatically separate narrow-line and broad-line AGN with high accuracy~\citep{Siudek2018, Siudek2018a, Siudek2022, Dubois2024}. Autoencoder architectures successfully recover AGN candidates missed by traditional BPT diagnostics due to low-SNR line measurements (Alcolea et al., in prep.), while diffusion-based models identify AGN from single-band morphological analysis of Euclid optical images~\citep{Stevens2025}. Foundation models may further unify multi-wavelength AGN detection, though their full potential remains to be explored~\citep{Siudek2025}. 
The multi-wavelength approach, such as the systematic incorporation of SPHEReX spectroscopic data into LSST AGN selection pipelines, analogous to the DESI Legacy Surveys and WISE approach presented here, could reduce the 15\% star-forming galaxy contamination observed in optical-only classifications. More broadly, a multi-modal approach that combines morphological, spectral, and photometric data may be key to robust AGN identification in the era of next-generation surveys.

\begin{acknowledgements} 
The authors thank the anonymous referee for insightful comments. 
This work has been supported by the Polish National Agency for Academic Exchange (Bekker grant BPN/BEK/2021/1/00298/DEC/1) and the State Research Agency of the Spanish Ministry of Science and Innovation (grant PGC2018-100852-A-I00 and PID2021-126838NB-I00). 
M.M. acknowledges support from the Spanish Ministry of Science and Innovation through the project PID2021-124243NB-C22, and the program Unidad de Excelencia Mar\'ia de Maeztu CEX2020-001058-M. H.Z. acknowledges the support from the National Natural Science Foundation of China (NSFC; grant Nos. 12120101003 and 12373010) and  National Key R\&D Program of China (grant Nos. 2023YFA1607800, 2022YFA1602902) and Strategic Priority Research Program of the Chinese Academy of Science (Grant Nos. XDB0550100).

This material is based upon work supported by the U.S. Department of Energy (DOE), Office of Science, Office of High-Energy Physics, under Contract No. DE–AC02–05CH11231, and by the National Energy Research Scientific Computing Center, a DOE Office of Science User Facility under the same contract. Additional support for DESI was provided by the U.S. National Science Foundation (NSF), Division of Astronomical Sciences under Contract No. AST-0950945 to the NSF’s National Optical-Infrared Astronomy Research Laboratory; the Science and Technology Facilities Council of the United Kingdom; the Gordon and Betty Moore Foundation; the Heising-Simons Foundation; the French Alternative Energies and Atomic Energy Commission (CEA); the National Council of Humanities, Science and Technology of Mexico (CONAHCYT); the Ministry of Science, Innovation and Universities of Spain (MICIU/AEI/10.13039/501100011033), and by the DESI Member Institutions: \url{https://www.desi.lbl.gov/collaborating-institutions}. Any opinions, findings, and conclusions or recommendations expressed in this material are those of the author(s) and do not necessarily reflect the views of the U. S. National Science Foundation, the U. S. Department of Energy, or any of the listed funding agencies.

The authors are honored to be permitted to conduct scientific research on Iolkam Du’ag (Kitt Peak), a mountain with particular significance to the Tohono O’odham Nation.

The DESI Legacy Imaging Surveys consist of three individual and complementary projects: the Dark Energy Camera Legacy Survey (DECaLS), the Beijing-Arizona Sky Survey (BASS), and the Mayall z-band Legacy Survey (MzLS). DECaLS, BASS and MzLS together include data obtained, respectively, at the Blanco telescope, Cerro Tololo Inter-American Observatory, NSF’s NOIRLab; the Bok telescope, Steward Observatory, University of Arizona; and the Mayall telescope, Kitt Peak National Observatory, NOIRLab. NOIRLab is operated by the Association of Universities for Research in Astronomy (AURA) under a cooperative agreement with the National Science Foundation. Pipeline processing and analyses of the data were supported by NOIRLab and the Lawrence Berkeley National Laboratory. Legacy Surveys also uses data products from the Near-Earth Object Wide-field Infrared Survey Explorer (NEOWISE), a project of the Jet Propulsion Laboratory/California Institute of Technology, funded by the National Aeronautics and Space Administration. Legacy Surveys was supported by: the Director, Office of Science, Office of High Energy Physics of the U.S. Department of Energy; the National Energy Research Scientific Computing Center, a DOE Office of Science User Facility; the U.S. National Science Foundation, Division of Astronomical Sciences; the National Astronomical Observatories of China, the Chinese Academy of Sciences and the Chinese National Natural Science Foundation. LBNL is managed by the Regents of the University of California under contract to the U.S. Department of Energy. The complete acknowledgments can be found at \url{https://www.legacysurvey.org/}.
\end{acknowledgements}

%%%%%%%%%%%%%%%%%%%%%%%%%%%%%%%%%%%%%%%%%%%%%%%%%%
\section*{Data Availability}

The VAC of physical properties of DESI EDR galaxies is publicly available at \url{https://data.desi.lbl.gov/doc/releases/edr/vac/cigale/}. 
The presented analysis is conducted on {\tt v1.4} based on the redshift coming from the {\tt QSO afterburner} pipeline relying on the {\tt QuasarNet} and the broad Mg II finder pipelines. 
The data behind the figures are available at \url{https://doi.org/10.5281/zenodo.15622223}. 

\bibliographystyle{aa}
\bibliography{DESIPhysPropCat}

%%%%%%%%%%%%%%%%%%%%%%%%%%%%%%%%%%%%%%%%%%%%%%%%%%

%%%%%%%%%%%%%%%%% APPENDICES %%%%%%%%%%%%%%%%%%%%%

\appendix

\section{SED Fitting Parameters}\label{app:CIGALETable}
 To ensure transparency and reproducibility, we provide in Table~\ref{tab:SEDParameters} the full set of input parameters and grid values used in the SED fitting process with {\tt CIGALE}. These parameters follow the configuration adopted in the DESI EDR VAC~\citep{Siudek2024}. As {\tt CIGALE}-based analyses are sensitive to the choice and priors of model parameters, we use a grid that balances physical coverage with computational feasibility.

\begin{table*}
\centering
        \caption{Default grid of input parameters used in SED fitting with {\tt CIGALE}. This configuration defines the model library used in this study and follows the configuration adopted in the DESI EDR VAC presented in~\citet{Siudek2024}.}
        \label{tab:SEDParameters}
        \footnotesize
        \begin{tabular}{r r r}
\hline
\hline
Parameter & Symbol &  Values\\
\hline
\multicolumn{3}{c}{Stellar population models:  \cite{Bruzual2003}}\\
\hline \hline
Initial mass function & IMF & \cite{Chabrier2003}\\
Metallicity & $\rm Z$ & 0.02\\
\hline
\multicolumn{3}{c}{SFH: Double exponentially decreasing}\\
\hline \hline
$\tau$ of the main stellar population (Gyr) & $\rm \tau_{main}$ & {0.1, 0.5, 1, 3, 5, 8}\\
Age of the main stellar population (Gyr) & $\rm t_1$ & 0.5, 1, 3, 4.5, 6, 8, 10, 13 \\
$\tau$ of the burst stellar population (Gyr) & $\rm \tau_{burst}$ & 10\\
Age of the burst stellar population (Gyr) & $\rm t_{burst}$ & 0.05\\
Mass fraction of young stellar population & $\rm f_{ySP}$ & 0, 0.01, 0.1, 0.2\\
\hline
\multicolumn{3}{c}{Nebular emission}\\
\hline \hline
Ionization parameter & $\rm logU$ & -2\\
Gas metallicity & $\rm Z_{gas}$ & 0.02\\

\hline
\multicolumn{3}{c}{Dust attenuation: \cite{Calzetti2000}}\\
\hline \hline
Color excess of the nebular emission & $\rm E(B-V)_{line}$ & 0, 0.05, 0.15, 0.3, 0.5, 0.75, 0.9, 1.1, 1.3, 1.6\\
Reduction factor to apply on $\rm E(B-V)_{line}$ & $\rm E(B-V)_{star}/E(B-V)_{line}$ & 0.44\\
\hline
\hline
\multicolumn{3}{c}{Dust emission: \cite{Draine2014}}\\
\hline \hline
Mass fraction of PAHs  & $\rm q_{PAH}$ &  0.47, 1.12, 2.5, 3.19\\
Minimum radiation field & $\rm U_{min}$ & 15\\
Power law slope of the radiation field & $\rm \alpha$ & 2.0\\
Fraction illuminated from $\rm U_{min}$ to $\rm U_{max}$& $\gamma$ & 0.02 \\
\hline
\multicolumn{3}{c}{AGN: \cite{Fritz2006}}\\
\hline \hline
The angle between the equatorial axis and line-of-sight & AGNPSY [deg] & 0.001, 20.100, 40.1, 70.100, 89.990 \\
Contribution of the AGN to the total LIR & AGNFRAC &  0, 0.01, 0.1, 0.3, 0.5, 0.7, 0.9 \\
\hline \hline
\end{tabular}
\end{table*}

\section{The SEDs for representative examples}\label{app:SEDExemplary}
\begin{figure*}
 	\centerline{\includegraphics[width=0.8\textwidth]{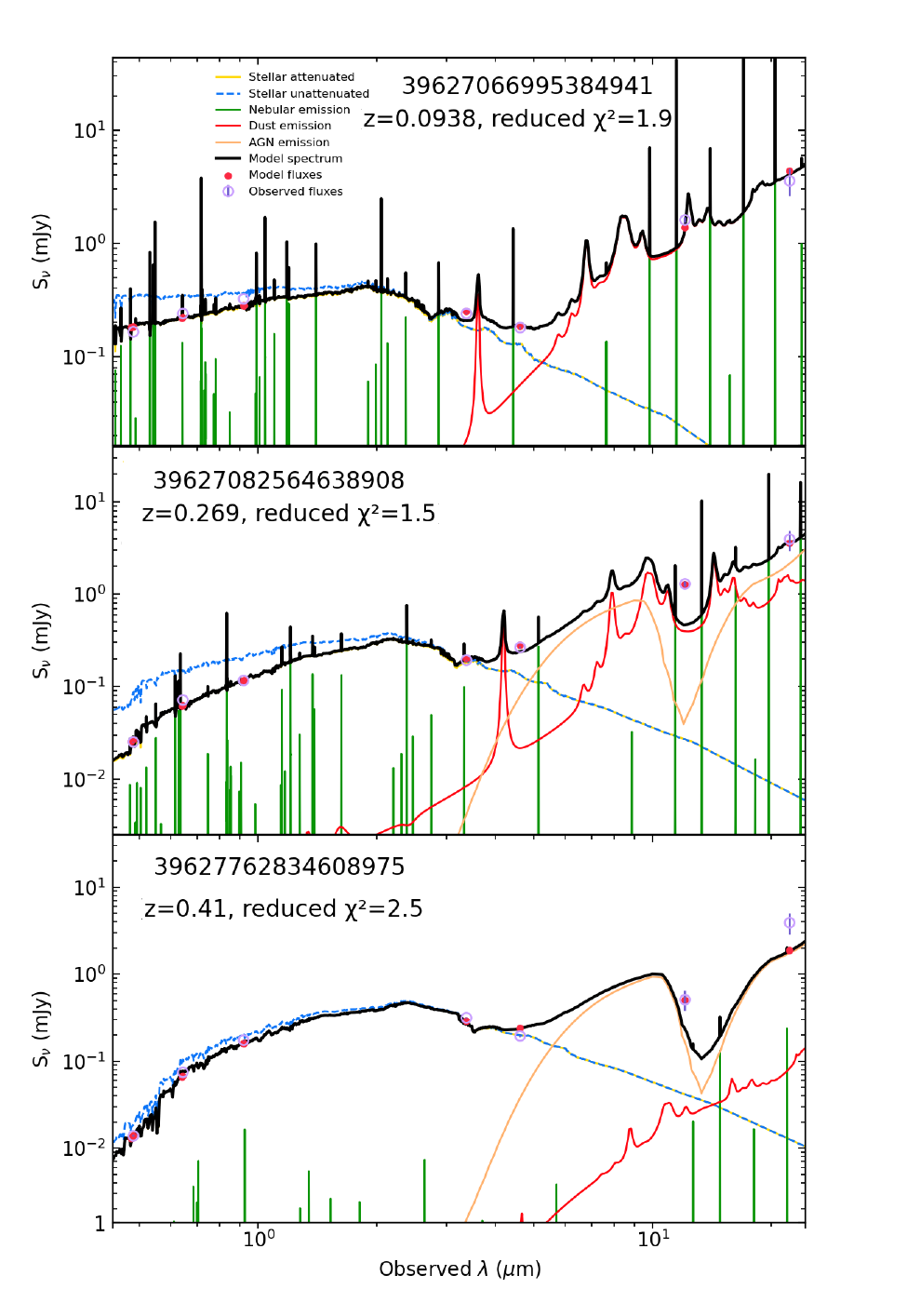}}
	\caption{Exemplary SEDs of star-forming galaxies (top), WISE/BPT-AGN (middle), and SED-AGN Only (bottom). The different curves represent the contributions from various components: stellar emission (attenuated and unattenuated), nebular emission, dust emission, and AGN emission. The black curve is the total model spectrum, and the observed fluxes are overlaid as violet points. The SED of a SED-AGN Only galaxy is characterized by a dominant AGN contribution in the MIR and a relatively weaker stellar component, suggesting a passive galaxy.  }
	\label{fig:SEDExemplary}
\end{figure*}

Figure~\ref{fig:SEDExemplary} shows the SEDs for representative examples from each category: a typical star-forming galaxy, a WISE/BPT-AGN, and a SED-AGN Only galaxy.
The star-forming galaxy exhibits a strong stellar component, with the SED peaking in the optical and NIR regions. The dust emission, indicated by the infrared excess, reflects ongoing star formation. The AGN contribution is negligible, consistent with the expected characteristics of a purely star-forming galaxy.
The WISE/BPT-AGN, in contrast, shows significant AGN activity, particularly in the MIR range. 
The SED reveals a combination of strong stellar emission and enhanced IR output due to AGN-heated dust. 
This suggests that while the galaxy is actively forming stars, the AGN is also contributing significantly to the overall energy output. 
The SED-AGN Only galaxy presents a distinct profile, with a relatively weaker stellar component and a dominant AGN contribution in the MIR. 
The optical part of the SED suggests a lower star formation rate or an older stellar population. The strong MIR excess is primarily driven by the AGN, indicating that these galaxies are likely passive, where AGN activity dominates.
These SED comparisons highlight the unique nature of SED-AGN Only galaxies, possibly representing a population of passive galaxies where the AGN is the dominant source of energy. 

\section{Dependence on model assumptions}\label{app:model_dependence}

When constructing the model grid for the SED fitting (see Sect.~\ref{sec:DesiEDR}), several assumptions are made regarding the SFH, metallicity, and dust attenuation laws. Here, we test the robustness of AGN identification against changes in these model assumptions using a representative sample of approximately 50\,000 galaxies from DESI EDR, covering a broad range of galaxy types (see Appendix D1 in~\cite{Siudek2024} for a description of the representative sample). 

Table~\ref{tab:model_dependence} summarizes the efficiency of AGN identification under different model choices relative to the default configuration. We always change only one parameter at a time, keeping all others fixed (e.g., when applying a non-parametric SFH of \cite{Ciesla2023}, we retain the IMF from \cite{Chabrier2003} and SSP models from \cite{Bruzual2003}). For a full description of our model components, see Sec. 6 in~\cite{Siudek2024}. We find the following:

\begin{itemize}
    \item Changes in the IMF to \cite{Salpeter1955} or the dust emission model to \cite{Dale2014} have negligible impact on the AGN selection efficiency.
    \item Adopting the \cite{CharlotFall2000} dust attenuation model reduces contamination by star-forming galaxies, but also decreases completeness for BPT-AGN and WISE-AGN.
    \item Using a non-parametric SFH~\citep{Ciesla2023} or the SSP models from \cite{Maraston2005} slightly increases the AGN identification rate, but at the cost of slightly increased contamination from star-forming galaxies.
    \item Allowing metallicity to vary as a free parameter improves both the completeness of AGN selection and reduces contamination.
    \item Switching the AGN emission model to SKIRTOR~\citep{Stalevski2012,Stalevski2016} enhances completeness, notably for broad-line AGN (up to 92\% recovery) and WISE-AGN (97\%), but at the cost of increased contamination from star-forming galaxies.
\end{itemize}

The choice of model ultimately depends on the scientific goals. Increased contamination from star-forming galaxies is generally undesirable in AGN-focused studies, as it reduces sample purity and can bias physical interpretations.  However, any model-driven enhancement or suppression of certain populations should be interpreted cautiously, as it may not reflect the true galaxy demographics.  Thus, model selection should reflect the intended use—whether to maximize AGN completeness (e.g., for demographic studies) or to ensure purity (e.g., for SED-based analyses). In the next Section, we further quantify the performance of the AGN identification depending on the model assumptions.

\begin{table*}[h!]
\centering
\begin{tabular}{lccccccccc}
\hline
Category & N & Default & IMF & Dust & Dust & SFH & Z & SSP & AGN \\
& &  &  & atten. & emission&  &  & model & model\\
 &  & \% & \% & \% & \% & \% & \% & \% & \% \\
\hline \hline
{\tt BL AGN} & 1\,033 & {\bf 78} & 79 & 75 & 75 & 80 & 83 & 81 & 92 \\
{\tt BPT-AGN} & 1\,479 & {\bf 65} & 66 & 55 & 65 & 70 & 67 & 75 & 80\\
{\tt BPT-SF} & 1\,901 & {\bf 21} & 22 & 15 & 27 & 24 & 16 & 30 & 27\\
{\tt WISE-AGN} & 2\,244 & {\bf 85} & 86 & 75 & 81 & 89 & 89 & 87 & 97\\
\hline\hline
{\tt NII-AGN} & 2\,215 & {\bf 64} & 66 & 53 & 65 & 70 & 66 & 75 & 78\\
{\tt NII-SF} & 2\,118 & {\bf 26} & 26 & 19 & 30 & 29 & 20 & 34 & 32\\
{\tt NII-LINER} & 655 & {\bf 47} & 47 & 43 & 52 & 48 & 50 & 62 & 56\\
{\tt NII-COMPOSITE} & 3\,543 & {\bf 29} & 30 & 20 & 36 & 33 & 26 & 43 & 40\\
{\tt SII-AGN} & 1\,473 & {\bf 64} & 65 & 55 & 66 & 69 & 67 & 75 & 79\\
{\tt SII-SF} & 6\,336 & {\bf 30} & 31 & 21 & 36 & 34 & 27 & 42 & 40\\
{\tt SII-LINER} & 360 & {\bf 44} & 44 & 48 & 46 & 45 & 52 & 62 & 54\\
{\tt OI-AGN} & 1\,700 & {\bf 65} & 66 & 56 & 65 & 70 & 68 & 75 & 80\\
{\tt OI-SF} & 6\,311 & {\bf 31} & 31 & 21 & 37 & 34 & 27 & 42 & 40\\
{\tt OI-LINER} & 333 & {\bf 48} & 48 & 50 & 46 & 48 & 52 & 63 & 57\\
{\tt WHAN-AGN} & 6\,301 & {\bf 40} & 40 & 29 & 45 & 44 & 38 & 52 & 51\\
{\tt WHAN-SF} & 2\,069 & {\bf 34} & 34 & 28 & 35 & 36 & 30 & 41 & 42\\
{\tt WHAN-RG} & 876 & {\bf 51} & 51 & 50 & 49 & 50 & 53 & 63 & 59\\
{\tt X-RAY-AGN} & 440 & {\bf 25} & 25 & 19 & 22 & 28 & 23 & 31 & 32\\
{\tt RADIO-AGN} & 1\,141 & {\bf 24} & 25 & 19 & 24 & 27 & 24 & 31 & 32\\
{\tt SED-AGN Only} & 4\,291 & {\bf 8} & 7 & 8 & 6 & 7 & 7 & 6 & 7\\
\hline
\end{tabular}
\caption{
Impact of model assumptions on the efficiency of SED-based AGN classification. Each row corresponds to a reference category (AGN, star-forming, LINERs, composite, and retired galaxies) based on standard selection methods (e.g., BPT, WISE, X-ray, radio). Values represent the percentage of objects in each category that are also classified as AGN by the SED-fitting approach under different model configurations (relative to the default setup; see text for details). For the row {\tt SED-AGN Only}, values indicate the fraction of galaxies identified solely by the SED method, not overlapping with any standard AGN selection. 
}

\label{tab:model_dependence}
\end{table*}

We additionally note that when WISE W3 and W4 bands are excluded from the SED fitting, the contamination from star-forming galaxies rises significantly, with recovery rates dropping for BPT-AGN, WISE-AGN, and broad-line AGN. In extreme cases where no MIR information is used, virtually all galaxies are classified as AGN ({\tt AGNFRAC}$\geq0.1$), illustrating the critical importance of MIR data in controlling contamination.

\section{AGN selection using {\tt AGNFRAC}}\label{app:ROC}

\begin{figure*}[h!]
    \centering
    \includegraphics[width=0.95\textwidth]{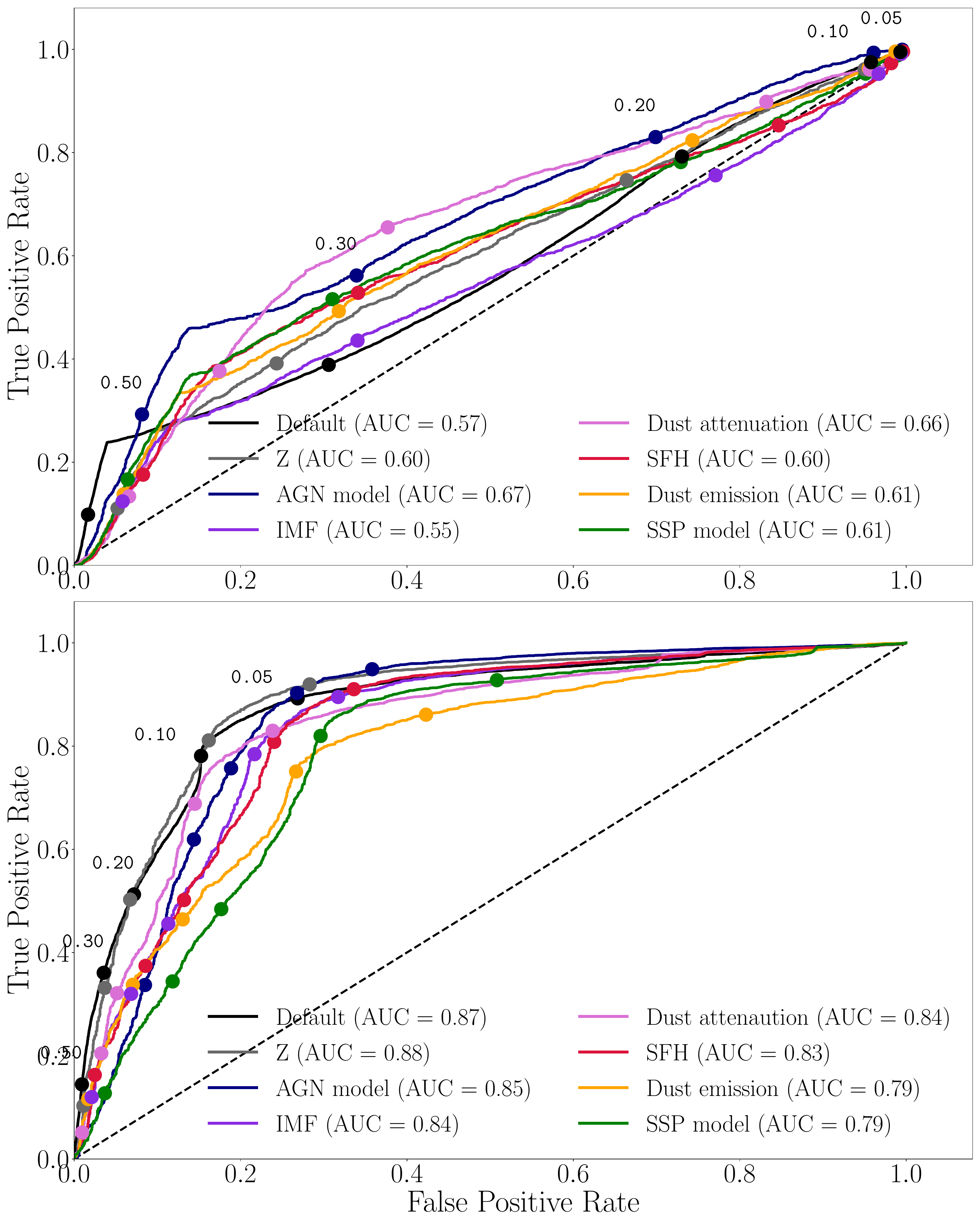}
    \caption{ROC curves for default {\tt CIGALE} configuration and different model choices (see App.~\ref{app:model_dependence}) showing the performance of {\tt AGNFRAC} as a classifier for AGN selection, comparing sources with poor-to-moderate infrared quality ({\tt FLAGINFRARED} $\leq 2$, left) and high-quality WISE photometry ({\tt FLAGINFRARED} $= 4$, right). Colored points indicate {\tt AGNFRAC} thresholds.}
    \label{fig:ROC}
\end{figure*}

 We assess the performance of the {\tt AGNFRAC} parameter, derived from SED fitting, as a tool for AGN classification by performing a ROC analysis for: i) sources with high-quality mid-IR photometry ({\tt FLAGINFRARED} $= 4$), and ii) those with poor or missing WISE coverage ({\tt FLAGINFRARED} $\leq 2$). A ROC curve is a representation of the true positive rate (TPR) versus the false positive rate (FPR) for different classification thresholds. The TPR and FPR are defined as:
\begin{equation}
    \mathrm{TPR} = \frac{\mathrm{TP}}{\mathrm{TP} + \mathrm{FN}}, \quad
    \mathrm{FPR} = \frac{\mathrm{FP}}{\mathrm{FP} + \mathrm{TN}} ,
\end{equation}
where TP, FP, FN, and TN are the counts of true positives (AGN that is classified as AGN), false positives (non-AGN that is classified as AGN), false negatives (AGN that is not classified as AGN), and true negatives (non-AGN that is not classified as AGN), respectively. A robust classification is achieved when AGN are classified as AGN ($\rm TPR = 1$) rather than AGN as non-AGN ($\rm FNR = 0$). 

As shown in Fig.~\ref{fig:ROC}, the ROC curve for the {\tt FLAGINFRARED} $= 4$ subset yields a significantly higher area under the curve ($\rm AUC = 0.87$\footnote{AUC score represents the probability that the classifier ranks a randomly chosen positive instance higher than a randomly chosen negative one. An AUC of 0.87, for example, indicates an 87\% chance of correctly distinguishing between AGN and non-AGN galaxies.}), indicating strong separability between AGN and star-forming galaxies using {\tt AGNFRAC}. In contrast, the ROC for the lower-quality subset ({\tt FLAGINFRARED} $\leq 2$) results in a much flatter curve with $\rm AUC = 0.57$, close to random classification. 
 Annotated thresholds on the ROC curves demonstrate that for {\tt FLAGINFRARED} $= 4$, completeness increases sharply with minimal contamination, especially around AGNFRAC values of $0.1 - 0.3$. In the lower-quality sample, no such threshold yields both high true positive and low false positive rates. 
 
 The ROC curve is sensitive to model assumptions (see App.~\ref{app:model_dependence}). For a subset with {\tt FLAGINFRARED} $= 4$ and metallicity left as a free parameter, the performance is similar or even slightly better ($\rm AUC = 0.88$) than for the default configuration (see also Tab.~\ref{tab:model_dependence}).

\end{document}